\documentclass{aa} 

\usepackage{graphicx}
\usepackage[varg]{txfonts}
\usepackage{color}
\usepackage[colorlinks=true,citecolor=blue,linkcolor=blue]{hyperref}
\usepackage{gensymb}
\usepackage{subcaption}
\usepackage{soul} 
\usepackage{amsmath} 

\graphicspath{{./figs/}}
\newcommand{\chandra}{\textsl{Chandra}\xspace}
\newcommand{\nustar}{\textsl{NuSTAR}\xspace}
\newcommand{\xmmnewton}{\textsl{XMM-Newton}\xspace}
\newcommand{\swift}{\textsl{Swift}\xspace}
\newcommand{\INTG}{\textsl{INTEGRAL}\xspace}
\newcommand{\xrism}{\textsl{XRISM}\xspace}
\newcommand{\athena}{\textsl{Athena}\xspace}

\newcommand{\angstrom}{\text{\normalfont\AA}}

\newcommand{\Lya}{Ly${\rm \alpha}$}
\newcommand{\Lyb}{Ly${\rm \beta}$}
\newcommand{\Lyg}{Ly${\rm \gamma}$}
\newcommand{\Hea}{He${\rm \alpha}$}
\newcommand{\OVIII}{O\,{\sc viii}}
\newcommand{\NeIX}{Ne\,{\sc ix}}
\newcommand{\NeX}{Ne\,{\sc x}}
\newcommand{\MgXI}{Mg\,{\sc xi}}
\newcommand{\MgXII}{Mg\,{\sc xii}}
\newcommand{\SiXIII}{Si\,{\sc xiii}}
\newcommand{\SiXIV}{Si\,{\sc xiv}}
\newcommand{\SXV}{S\,{\sc xv}}
\newcommand{\SXVI}{S\,{\sc xvi}}
\newcommand{\ArVIIX}{Ar\,{\sc vi--ix}}
\newcommand{\ArXVII}{Ar\,{\sc xvii}}
\newcommand{\ArXVIII}{Ar\,{\sc xviii}}

\newcommand{\CaIIXII}{Ca\,{\sc ii--xii}}

\usepackage{footmisc}
\usepackage[dvipsnames]{xcolor}

\begin{document}

\title{Observing the onset of the accretion wake in Vela X-1}

\author{
C.~M.~Diez\inst{1}\and
V.~Grinberg\inst{2}\and
F.~F\"urst\inst{3} \and
I.~El~Mellah\inst{4}\and
M.~Zhou\inst{1}\and
A.~Santangelo\inst{1}\and
S.~Mart\'inez-N\'u\~nez\inst{5}\and
R.~Amato\inst{6}\and
N.~Hell\inst{7}\and
P.~Kretschmar\inst{8}
}

\institute{Institut f\"ur Astronomie und Astrophysik, Universität T\"ubingen, Sand 1, 72076 T\"ubingen, Germany \\
\texttt{diez@astro.uni-tuebingen.de}
\and
European Space Agency (ESA), European Space Research and Technology Centre (ESTEC), Keplerlaan 1, 2201 AZ Noordwijk, The Netherlands
\and
Quasar Science Resources S.L for European Space Agency (ESA), European Space Astronomy Centre (ESAC), Camino Bajo del Castillo s/n, 28692 Villanueva de la Cañada, Madrid, Spain
\and 
Université Grenoble Alpes, CNRS, IPAG, Grenoble, France
\and
Instituto de F\'isica de Cantabria (CSIC-Universidad de Cantabria), E-39005, Santander, Spain
\and
IRAP, CNRS, Université de Toulouse, CNES, 9 Avenue du Colonel Roche, 31028 Toulouse, France
\and
Lawrence Livermore National Laboratory, 7000 East Avenue, Livermore CA 94550, USA
\and
European Space Agency (ESA), European Space Astronomy Centre (ESAC), Camino Bajo del Castillo s/n, 28692 Villanueva de la Cañada, Madrid, Spain  
}

\date{Received 16 December 2022/ Accepted 16 March 2023}

\abstract{High-Mass X-ray Binaries (HMXBs) offer a unique opportunity for the investigation of accretion onto compact objects and of wind structure in massive stars. A key source for such studies is the bright neutron star HMXB Vela~X-1 whose convenient physical and orbital parameters facilitate the analysis and in particular enable studies of the wind structure in HMXBs. Here, we analyse simultaneous \xmmnewton and \nustar observations at $\phi_{\mathrm{orb}} \approx$ 0.36--0.52 and perform time-resolved spectral analysis down to the pulse period of the neutron star, based on our previous \nustar-only results. For the first time, we are able to trace the onset of the wakes in a broad 0.5--78\,keV range with a high-time resolution of $\sim$283\,s and compare to theoretical predictions. We observe a clear rise of the absorption column density of the stellar wind $N_{\mathrm{H,1}}$ starting at orbital phase $\sim$0.44, corresponding to the wake structure entering our line of sight towards the neutron star, together with local extrema throughout the observation possibly associated with clumps or other structures in the wind. Periods of high absorption reveal the presence of multiple fluorescent emission lines of highly ionised species, mainly in the soft X-ray band between 0.5 and 4\,keV, indicating photoionisation of the wind.
}

   \keywords{X-rays: binaries,  stars: neutron,  stars: winds, outflows}

   \maketitle

\section{Introduction}
\label{section:intro}

The prototypical eclipsing HMXB \object{Vela X-1} is one of the brightest persistent point sources in the X-ray sky despite a moderate mean intrinsic luminosity of $5\times10^{36}$\,erg\,s$^{-1}$ \citep{Fuerst_2010a}. It lies at the relatively close distance of $1.99^{+0.13}_{-0.11}$~kpc. See the review by \citet{Kretschmar_2021a} and references therein for this and other system parameters quoted in the introduction. The system consists of a B0.5 Ib supergiant, \object{HD~77581} \citep{Hiltner_1972} orbited by an accreting neutron star with an orbital period of $\sim$8.964\,d \citep{Kreykenbohm_2008,Falanga:2015}. 
The mass and radius of HD~77581 have been estimated by different authors as between $\sim$20 and $\sim$26\,$M_{\odot}$ and between $\sim$27 and $\sim$32\,$R_{\odot}$ \citep[Table 4 in][]{Kretschmar_2021a}. A strong stellar wind with a mass loss rate of $\sim$10$^{-6}M_{\odot}\,$yr$^{-1}$ \citep[e.g.][]{Watanabe_2006, Falanga:2015, Gimenez-Garcia_2016} fuels the accretion of matter onto the neutron star and the pulsating X-ray emission with a fluctuating pulse period of $\sim$283\,s. The mass of the neutron star is estimated to be $\sim$1.7--2.1 $M_{\odot}$, thus on the heavier side of the typical mass distribution, see Table~4 in \citet{Kretschmar_2021a}, especially the estimates by \citet{Rawls_2011} and \citet{Falanga:2015}.

Since the orbital separation of the two components is only $\sim$1.7\,$R_{\star}$ \citep{vanKerkwijk_1995, Quaintrell_2003}, the neutron star is deeply embedded in the dense wind of the supergiant and significantly influences the wind flow. Due to the high inclination of the system of $>73\degree$ \citep[and others]{JossRappaport:1984,vanKerkwijk_1995}, observations at different orbital phases can sample different structures in the stellar wind as modified by the interaction with the neutron star. One notable feature is the varying absorption as function of orbital phase, found by many authors \citep[see the overview Fig.~5 of][]{Kretschmar_2021a}, with typically a decrease of the absorption after eclipse up to an orbital phase of 0.2--0.3 \citep{Martinez_2014, Lewis_1992} followed by a sometimes steep increase in the phase range 0.4--0.6 \citep{Haberl_1990}, with a varying location in phase in individual observations and with generally high absorption at later orbital phases \citep{Sato_1986, Haberl_1990}. 

In addition to the variations in observed flux due to X-ray absorption, wind accreting X-ray pulsars also usually show significant variability of the intrinsic X-ray flux caused by a complex interplay of factors like variations in the density of the accreted material, possible inhibition of accretion by interactions at the magnetosphere, etc. \citep[][]{Martinez_2017}. For individual observations, especially at softer X-ray energies, it is not always evident to distinguish absorption variations from those of the intrinsic emission. A study of variability in Vela~X-1 by \citet{Fuerst_2010a} based on \INTG data in the hard X-ray band, which is only very little affected by absorption, found a log-normal distribution of the intrinsic flux values. On various occasions 'low states' or 'off-states' have been observed \citep[see Table~1 in][with multiple references]{Kretschmar_2021a}, but again not always with a clear distinction of intrinsic X-ray brightness versus high absorption.

Already early X-ray observations and optical emission line evidence suggested a wake structure trailing the X-ray source in the binary system \citep[see references in][]{Conti_1978}. These structures can be caused by multiple processes.
The neutron star moving highly supersonically through the dense stellar wind will lead to a bow shock with a trailing accretion wake. 
The photoionisation of the wind by the bright X-ray source can lead to the formation of a shock between the accelerating wind and the stalling photoionised plasma \citep{Fransson+Fabian_1980}, which then can lead to a trailing spiral structure, the photoionisation wake. X-ray heating and radiative cooling of the wind \citep{Kallman+McCray_1982,Blondin_1990a} can create additional instabilities and filamentary structures, leading to stronger short-term variations. 
Tidal deformation of the mass donor can lead to enhanced wind flux in the direction of the neutron star, even if it does not fill its Roche Lobe. For close systems this can develop into a dense 'tidal stream' \citep{Blondin_1991}, which deflected by Coriolis force tends to pass behind the compact object and thus explains high absorption at later orbital phases. The density enhancement from such a tidal stream would be expected to be quite stable in orbital phase, while variations caused by an accretion wake would vary from orbit to orbit. According to \citet{Blondin_1994AIPC}, in a specific system there would be either a tidal stream between the star and the neutron star, or a photoionisation wake trailing the neutron star.

Vela~X-1 has also been studied via X-ray line spectroscopy. Early studies \citep[e.g.][]{Becker_1978,Ohashi_1984} mainly focused on the pronounced iron line complex visible also very clearly in eclipse. A broad FeK$\alpha$ emission line was also found by \citet{Sato_1986}, but they suspected further contributions to the overall line intensity from FeK$\beta$ as well as from fluorescent K$\alpha$ lines of Si, S, Ar, Ca, and Ni. Further elements have been reported from spectra taken during, or close to eclipse \citep{Nagase_1994,Sako_1999a,Schulz_2002}, including recombination lines and Radiative Recombination Continua (RRC), as well as fluorescence lines. The variety of ionisation states observed strongly indicates the presence of an inhomogeneous wind with optically thick, less ionised matter coexisting with warm photoionised plasma. \chandra/HETGS spectra from three orbital phases (0, 0.25 and 0.5) were analysed both by \citet{Goldstein_2004} and \citet{Watanabe_2006}, finding an eclipse-like spectrum around orbital phase 0.5, while around phase 0.25 the spectrum was dominated by the continuum. \citet{Grinberg_2017} revisited the observation at phase 0.25 analysing time intervals of low and high spectral hardness (i.e. different levels of absorption) separately and detected line features from high and low ionisation species of Si, Mg, and Ne, as well as strongly variable absorption. This again implied the presence of both cool and hot gas phases, possibly from the combination of an intrinsically clumpy stellar wind and a highly structured accretion flow close to the compact object. Analysing the grating spectrometer data of a long \xmmnewton observation at early orbital phases, \citet{Lomaeva_2020a} found emission lines corresponding to highly ionised O, Ne, Mg, and Si as well as RRC of O. In addition, they found potential absorption lines of Mg at a lower ionisation stage and features identified as iron L lines. 

In our targeted observing programme with \xmmnewton and \nustar, we aimed to cover an interesting binary phase range in which we expect strong changes in absorption while the accretion and ionisation wakes are crossing the line of sight of the observer \citep{Grinberg_2017}. In our previous work \citep{Diez_2022} with \nustar, we observed strong absorption variability along the orbital phase. However, an in-depth study of the varying absorption cannot be explored with \nustar alone as we need coverage at lower energies, as possible with \xmmnewton. On the other hand, \xmmnewton alone is not sufficient and we need \nustar to constrain the continuum. Hence, this is why a simultaneous observation was necessary.

We introduced the datasets and their reduction in Sect.~\ref{section:data_reduc} followed by the presentation of the light curves extracted in different relevant energy bands in Sect.~\ref{section:lightcurves}. Using the results from the latter section, we proceed to the time-resolved spectral analysis in Sect.~\ref{section:spectrum_analysis} and finally discuss our results in Sect.~\ref{section:discussion} focusing on the absorption variability. A summary is given in Sect.~\ref{section:summary} and an Appendix~\ref{appendix:calib_title} is given to describe in detail the calibration issues we faced during this work between \nustar and \xmmnewton.

\section{Observation and data reduction}
\label{section:data_reduc}
\begin{figure}
    \centering
    \centerline{\includegraphics[trim=0cm 15cm 9.5cm 14cm, clip=true, width=1\linewidth]{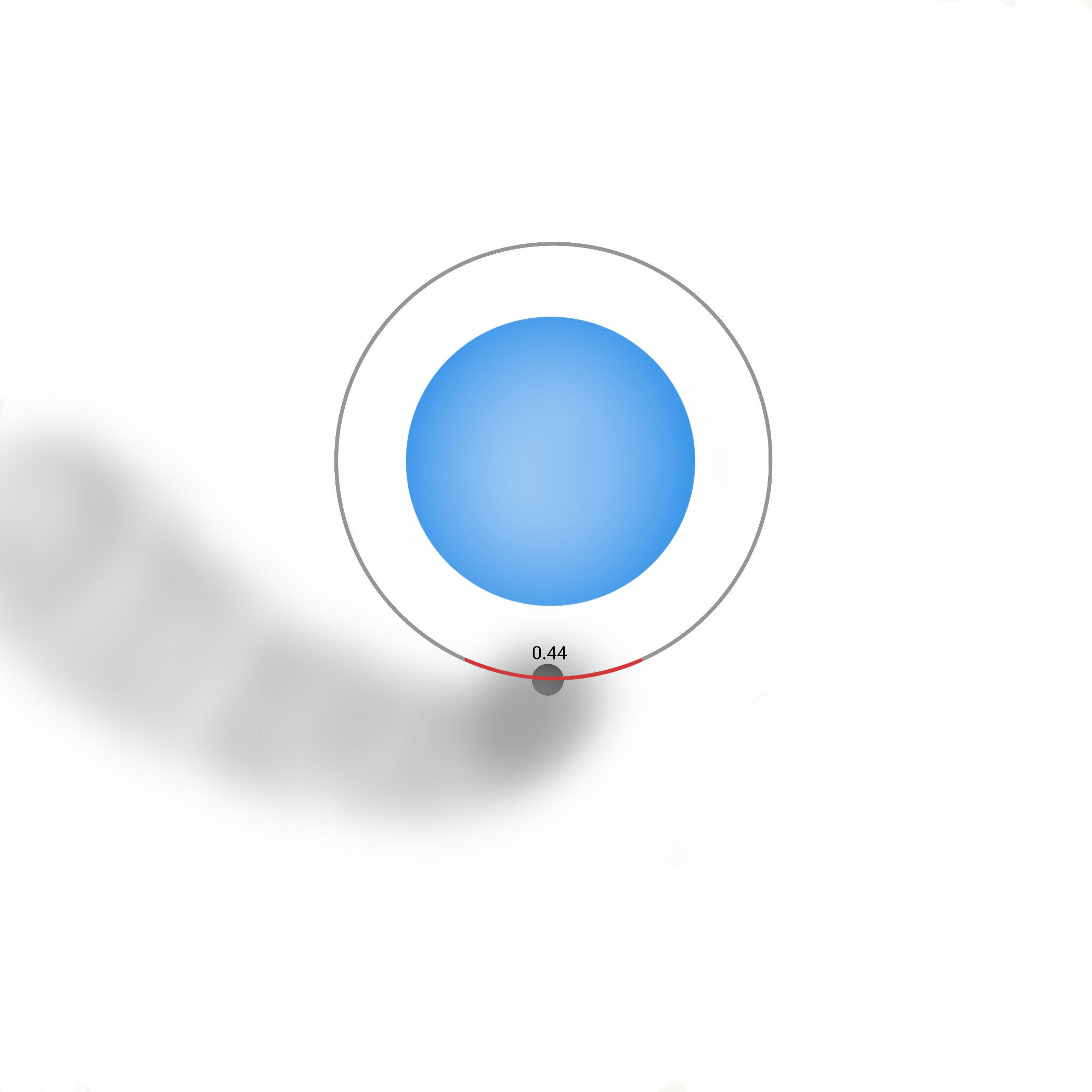}}
    \caption{Sketch of Vela X-1 showing the orbital phases covered during this observation. In this image, the observer is located facing the system at the bottom of the picture.}
    \label{fig:vela_sketch}
\end{figure}

Vela X-1 was observed on 3--5 May 2019 as science target of a simultaneous campaign with \xmmnewton and \nustar. Onboard \xmmnewton \citep{Jansen_2001} under Obs ID 0841890201, the European Photon Imaging Camera pn-CCDs \citep[EPIC-pn;][]{Strueder_2001}, the EPIC Metal Oxide Semi-conductor \citep[EPIC-MOS;][]{Turner_2001} and the Reflection Grating Spectrometers \citep[RGS;][]{denHerder_2001} were used. The \nustar data under Obs ID 30501003002 have already been analysed in \citet{Diez_2022}. In this paper, we focus on the simultaneous analysis of the new EPIC-pn data with our previous \nustar work (updated with the current calibration files and software) in order to have a broadband mapping of the stellar wind along the orbital phase of the neutron star around the companion star. The simultaneous RGS data will be explored in a later work. Details about the observation are given in Table~\ref{tab:obs_details} and a sketch of the system during this observation is shown in Fig.~\ref{fig:vela_sketch}. We note the much larger net \xmmnewton EPIC-pn exposure due to the Low Earth Orbit (LEO) of \nustar.

The orbital phases $\phi_{\mathrm{orb}}$ are obtained with the ephemeris in Table~2 from \citet{Diez_2022} which is derived from \citet{Kreykenbohm_2008} and \citet{Bildsten_1997} and where $\phi_{\mathrm{orb}} = 0$ is defined as $T_{90}$. The time of phase zero is usually defined by $T_{90}$ (the time when the mean longitude $l$ is equal to 90\degree) or by $T_{\rm{ecl}}$ (the mid-eclipse time). Explanations on how to convert $T_{90}$ to $T_{\rm{ecl}}$ can be found in \citet{Kreykenbohm_2008}. In Table~\ref{tab:obs_details}, orbital phases with $T_{\mathrm{ecl}}$ are also given. In this work, we exclusively use $T_{90}$ and all the mentioned times are corrected for the binary orbit.

We use HEASOFTv6.30.1. To analyse the data, we use the Interactive Spectral Interpretation System (ISIS) v1.6.2-51 \citep{Houck&Denicola_2000}. ISIS provides access to XSPEC \citep{Arnaud_1996a} models that are referenced later in the text.

\begin{table*}
\renewcommand{\arraystretch}{1.1}
\caption{Observations log.}
\label{tab:obs_details}
\begin{center}
\begin{small}
\begin{tabular}{lcccccc}    
\hline\hline
Instrument &
Obs ID & 
\multicolumn{1}{c}{Time Start} & 
\multicolumn{1}{c}{Time Stop} &
\multicolumn{1}{c}{Exposure} &
\multicolumn{1}{c}{Orbital phase} &
\multicolumn{1}{c}{Orbital phase} \\
&
&
\multicolumn{1}{c}{MJD (day) binarycor$^{(a)}$} & 
\multicolumn{1}{c}{MJD (day) binarycor$^{(a)}$} &
\multicolumn{1}{c}{(ks)} &
\multicolumn{1}{c}{(with $T_{\mathrm{90}}$)} &
\multicolumn{1}{c}{(with $T_{\mathrm{ecl}}$)} \\
\hline
\multicolumn{1}{l} \nustar & {30501003002}     & 58606.8688     & 58608.2465     & 40.562       & 0.36--0.52     & 0.34--0.49       \\
\multicolumn{1}{l} \xmmnewton & {0841890201}     & 58606.9283     & 58608.2146     & 109.311       & 0.37--0.51     & 0.34--0.48       \\
\hline
\multicolumn{7}{p{0.9\linewidth}}{$^{(a)}$ Stands for 'binary corrected' times. The \texttt{BinaryCor} tool in ISIS removes the influence of the double-star motion for circular or elliptical orbits. The Git repository of the \texttt{isisscripts} where the \texttt{BinaryCor} function is described can be found at \url{https://www.sternwarte.uni-erlangen.de/gitlab/remeis/isisscripts/-/blame/6216a4ab8307f5825e17109db3cfb5c317a7ab08/share/isisscripts.sl}. Time start and stop are given in Modified Julian Dates (MJD).}\\
\end{tabular}
\end{small}
\end{center}
\renewcommand{\arraystretch}{1.0}
\end{table*}

\subsection{\nustar}
\label{section:nustar_evt_generation}

We used the NuSTARDAS pipeline (\texttt{nupipeline}) v2.1.2 with the calibration database (CALDB) v20220608 applied with the clock correction. 

We proceeded to the extraction of the products as in \citet{Diez_2022} with the updated pipeline and calibration files mentioned above. To summarise, we extracted a spectrum for every orbit of \nustar around the Earth and for every rotation of the neutron star with the previously derived pulse period of $\sim$283\,s. For the extraction of the source region, we used this time a smaller radius of $\sim$60 arcsec to minimise the impact of the background. The uncertainties are given at 90\% confidence and the events were barycentred using the \texttt{barycorr} tool from NUSTARDAS pipeline. The spectra were re-binned within ISIS to a minimal signal to noise of 5, adding at least 2, 3, 5, 8, 16, 18, 48, 72, and 48 channels for energies between 3.0–10, 10–15, 15–20, 20–35, 35–45, 45–55, 55–65, 65–76, and 76–79 keV, respectively, as in our previous work.

\subsection{\xmmnewton}

The observation was set-up in pn-timing mode with a thin filter. For the generation of the event lists file, we used the Science Analysis System (SAS) software v20.0 with the Current Calibration Files (CCF) as of April 2022 starting from Observation Data Files (ODFs) level running \texttt{epproc}\footnote{\label{footnote:epproc}\url{https://xmm-tools.cosmos.esa.int/external/sas/current/doc/epproc/node8.html}}. 

The default calibration uses \texttt{withrdpha='Y'} and \texttt{runepfast='N'} as of SASv14.0\footnote{see the \xmmnewton CCF Release Note 0369 (Migliari S., Smith M., 2019, XMM-CAL-SRN-0369) \url{https://xmmweb.esac.esa.int/docs/documents/CAL-SRN-0369-0-0.pdf}}. However, this default timing mode calibration lead us to an offset of the instrumental and physical lines of the source towards higher energies of $\sim$+140\,eV (see Appendix~\ref{appendix:iron_line_calib_epproc}). Therefore, after consulting with the \xmmnewton calibration team (S. Migliari, priv. comm.), we decided to turn off the Rate Dependent PHA (RDPHA) correction (\texttt{withrdpha='N'}) and apply the Rate Dependent CTI (RDCTI) correction, using \texttt{epfast} (\texttt{runepfast='Y'}), instead, resulting in satisfactory spectra.

No filtering for flaring particle background was necessary. Because the source was so bright that it illuminated the whole CCD detector, no background was extracted. The event times were barycentred using the \texttt{barycen} tool from the SAS pipeline and were deleted if on bad pixels with the \texttt{\#XMMEA\_EP} argument in the \texttt{evselect} step.

Since \xmmnewton EPIC-pn and \nustar observations were simultaneous, we could reuse the pulse period previously derived on the FPMA light curve, $P = 283.4447 \pm 0.0004\, \mathrm{s}$ \citep{Diez_2022}, to extract the \xmmnewton light curves to avoid intrinsic pulse variability. For a sanity check, we still performed epoch-folding on the \xmmnewton light curve with 1 sec binning and we indeed derived the exact same pulse period as in our previous work.

We selected single and double pixel events (\texttt{PATTERN<=4} in the \texttt{evselect} step) in the source region from \texttt{RAWX=32} to \texttt{RAWX=44}. We removed the outermost parts of the Point Spread Function (PSF) wings to reduce the influence of background noise or possible dust scattering effects. The count rate of the overall observation on different energy bands was high enough to select a region of only 13 pixels centred around the maximum of the PSF. We applied the task \texttt{epiclccorr} to perform absolute and relative corrections.

\subsubsection{Spectrum extraction and pile-up}
For the extraction of the spectra, we did the same selection of events than above but we also deleted events close to CCDs gaps or bad pixels (\texttt{FLAG==0} in the \texttt{evselect} step). An update to the EPIC effective area is available since April 2022 to improve the cross-calibration between \xmmnewton and \nustar as of April 5th 2022\footnote{\label{footnote:caltn0230} see the \xmmnewton Science Operations Team Calibration Technical Note 0230 (Fürst F., 2022, XMM-SOC-CAL-TN-0230) \url{https://xmmweb.esac.esa.int/docs/documents/CAL-TN-0230-1-3.pdf}}. As we want to compare our \xmmnewton results to our previous \nustar ones, we activated this correction using the \texttt{applyabsfluxcorr=yes} argument in the \texttt{arfgen} step. All spectra were re-binned using the optimal rebinning approach of \citet{Kaastra_2016}.

Because of the high count rate (see Sect.~\ref{section:lightcurves}), the data were strongly affected by pile-up \citep{Jethwa_2015} in particular at the centre of the PSF. To test and evaluate for pile-up in our observation, we used the \texttt{epatplot} task to read the pattern information statistics of the input EPIC-pn set as function of PI channel. As an additional sanity check, we performed an energy test on the iron line region (for more details, see Appendix~\ref{appendix:iron_line_pile_up}). 
According to the results of the aforementioned tests, we decided to exclude the three centremost columns of the PSF (\texttt{RAWX 37--38--39}). 

Moreover, we excluded spectral channels below 0.5\,keV and above 10\,keV from the analysis where the signal to noise ratio was very low.

We report in this work some cross-calibration issues between \nustar and \xmmnewton EPIC-pn. Cross-calibration issues between \xmmnewton EPIC-pn and \nustar are a known phenomenon when combining datasets from both observatories\footref{footnote:caltn0230} \citep[see e.g.][]{Gokus_master} and the \xmmnewton Science Operations Centre is working on this issue with a coming update of the SAS software (S. Migliari, private conversation). At the time of writing this work, no release was available so we described the cross-instrumental issues in Appendix~\ref{appendix:cross-instrumental_issues} and how we coped with them for our analysis in Sect.~\ref{subsection:spectral_analysis_orbit_XMM_NuSTAR}.

\section{Light curves and timing}
\label{section:lightcurves}

\subsection{Pulse period and average light curves}
\label{section:pp_lc}

Strong flux variability has been detected in Vela X-1 during this observation with timescales ranging from several kilo-seconds down to the pulse period of the neutron star. 

\begin{figure*}
    \centering
    \centerline{\includegraphics[trim=0cm 0cm 0cm 0.5cm, clip=true, width=1.0\linewidth]{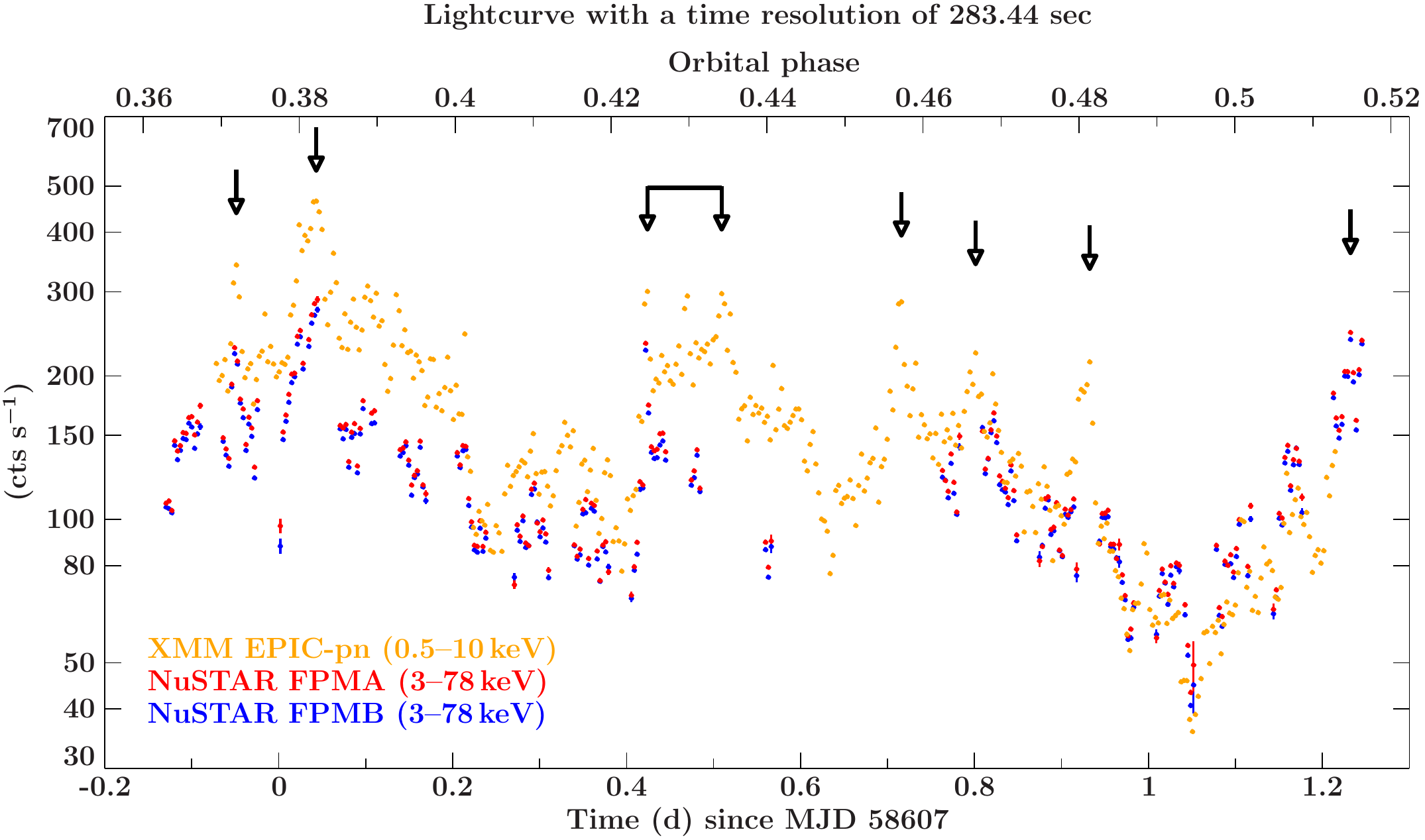}}
    \caption{Light curves for \xmmnewton EPIC-pn (orange), \nustar FPMA (red) and FPMB (blue) with a time resolution of $P = 283.44 \, \rm{s}$. The count rate is plotted against the orbital phase (top axis) and the time of the observation (bottom axis), y-axis in logarithmic scale. Major short flares are indicated by single arrows. The connected arrows at $\sim$58607.42\,$\mathrm{MJD}$ indicate a flaring episode of $\sim$8\,ks.}
    \label{fig:both_lc}
\end{figure*}

To study the overall system behaviour, we present in Fig.~\ref{fig:both_lc} the $0.5$--$10\,\mathrm{keV}$ \xmmnewton EPIC-pn light curve together with the \nustar FPMA and FPMB $3$~--$78\,\mathrm{keV}$ light curves. 

The ratio of the \xmmnewton EPIC-pn count rate by the \nustar FPMA-FPMB count rate is not constant. During the first half of the observation (until $T_{\mathrm{obs}} \approx 58607.60 \ \mathrm{MJD}$), the average \xmmnewton EPIC-pn count rate is $\sim$30\% higher than the \nustar one. Their ratio stabilises around 1 during the second half of the observation when the wakes are coming through our line of sight (see Fig.~\ref{fig:vela_sketch}) and therefore when the absorption from the stellar wind is more prominent. Since \xmmnewton EPIC-pn covers lower energies than \nustar, we expect it to be more affected by the absorption due to the stellar wind, explaining this behaviour.

All major flares detected are indicated by arrows in Fig.\ref{fig:both_lc}.  We can observe three flares happening simultaneously with both \nustar and \xmmnewton EPIC-pn  at $T_{\mathrm{obs}} \approx 58606.95 \ \mathrm{MJD}$, $T_{\mathrm{obs}} \approx 58607.04 \ \mathrm{MJD}$ and $T_{\mathrm{obs}} \approx 58607.42 \ \mathrm{MJD}$ and a fourth flare visible at $T_{\mathrm{obs}} \approx 58608.23 \ \mathrm{MJD}$ covered by \nustar only, as expected in \citet{Diez_2022}. With the new addition of EPIC-pn data, we can retrieve the data between $T_{\mathrm{obs}} \approx 58607.57 \ \mathrm{MJD}$ and $T_{\mathrm{obs}} \approx 58607.76 \ \mathrm{MJD}$ which were lost during the \nustar campaign and also retrieve data during the eclipses of the \nustar instrument. Therefore, we can observe three new flares at $T_{\mathrm{obs}} \approx 58607.72 \ \mathrm{MJD}$, $T_{\mathrm{obs}} \approx 58607.80 \ \mathrm{MJD}$ and $T_{\mathrm{obs}} \approx 58607.93 \ \mathrm{MJD}$ together with the flare at $T_{\mathrm{obs}} \approx 58607.42 \ \mathrm{MJD}$ being longer than what has been seen with \nustar. This flare lasts $\sim$8\,ks and reaches $\sim$300\,counts$\,$s$^{-1}$ which is almost as long but brighter than the flaring period in \citet{Martinez_2014}. The timescales of the flares appear to be from less than a \nustar eclipse ($\sim$2.5\,ks) up to $\sim$8\,ks. The brightest observed flare at $T_{\mathrm{obs}} \approx 58607.04 \ \mathrm{MJD}$ reaches $\sim$465\,counts$\,$s$^{-1}$ with EPIC-pn.

\subsection{Energy-resolved light curves}

\begin{figure}
    \centering
    \centerline{\includegraphics[trim=0cm 0cm 0cm 0cm, clip=true, width=1.0\linewidth]{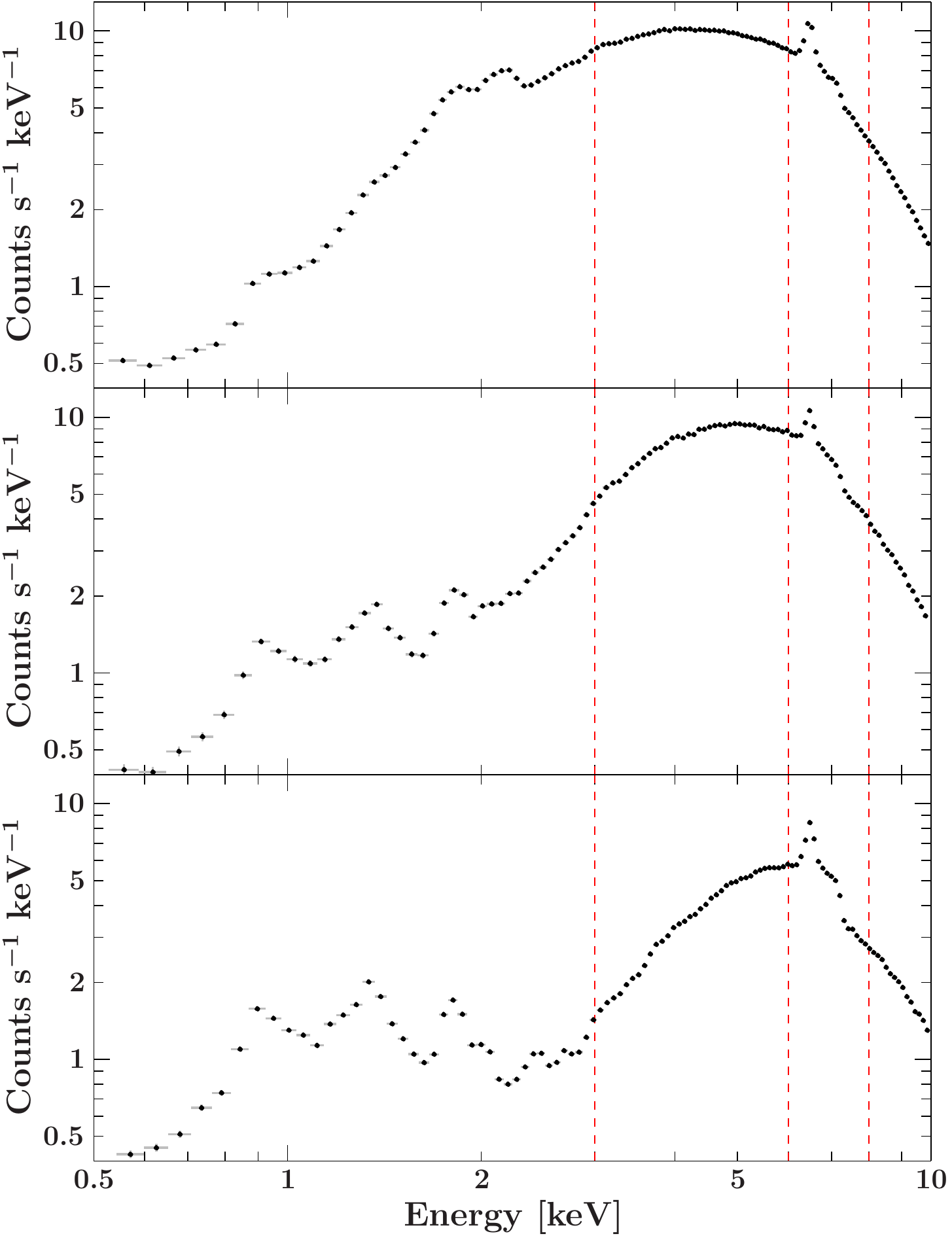}}
    \caption{\xmmnewton EPIC-pn spectra extracted during the three different phases of the observation in chronological order from top to bottom panel: phase of stable hardness ratio ($0.37 \lesssim \phi_{\mathrm{orb}} \lesssim 0.44$), phase of the loss of \nustar data ($0.44 \lesssim \phi_{\mathrm{orb}} \lesssim 0.46$) and phase of the rise of the hardness ratio ($0.46 \lesssim \phi_{\mathrm{orb}} \lesssim 0.51$), respectively. The vertical red dashed lines indicate the four energy bands we chose for the extraction of the energy-resolved light curves.}
    \label{fig:evol_spectra}
\end{figure}

To estimate the influence of the stellar wind absorption on the observed count rate variations, we extract the light curves in relevant energy bands. 
In \citet{Diez_2022}, we saw a change in the hardness ratio between the 3.0--5.0\,keV and 20.0--30.0\,keV energy bands roughly separating the observation into three noticeable phases: stable hardness ratio from the beginning of the observation to $T_{\mathrm{obs}} \approx 58607.57 \ \mathrm{MJD}$, followed by the loss of the \nustar data until $T_{\mathrm{obs}} \approx 58607.76 \ \mathrm{MJD}$ and finally the rise of the hardness ratio until the end. We present in Fig.~\ref{fig:evol_spectra} three different \xmmnewton EPIC-pn spectra taken during the above mentioned phases of the observation. We can observe low-energy variability towards the end of the observation (last panel of the figure). As expected from the geometry of the system (see Fig.~\ref{fig:vela_sketch}), when the wakes occupy most of the line of sight of the observer the absorption from the wind is so strong that the emission lines of the material become visible \citep[see e.g.][]{Watanabe_2006}. From the last spectrum, we can even highlight four energy bands of interest. The first one from 0.5 keV to 3.0 keV covers all the low-energy emission lines. We choose the second energy band from 3.0 keV to 6.0 keV to account for the low-energy part of the continuum before the iron line region from 6.0 keV to 8.0 keV. The last energy band covers the high-energy part of the continuum from 8.0 keV to 10.0 keV. This will help us comparing the photon count rate in different energy bands relatively to the continuum to check for variability. We will present a more detailed spectral analysis in Sect.~\ref{section:spectrum_analysis}.

\begin{figure*}
    \centering
    \centerline{\includegraphics[trim=0cm 0cm 0cm 0.5cm, clip=true, width=1.0\linewidth]{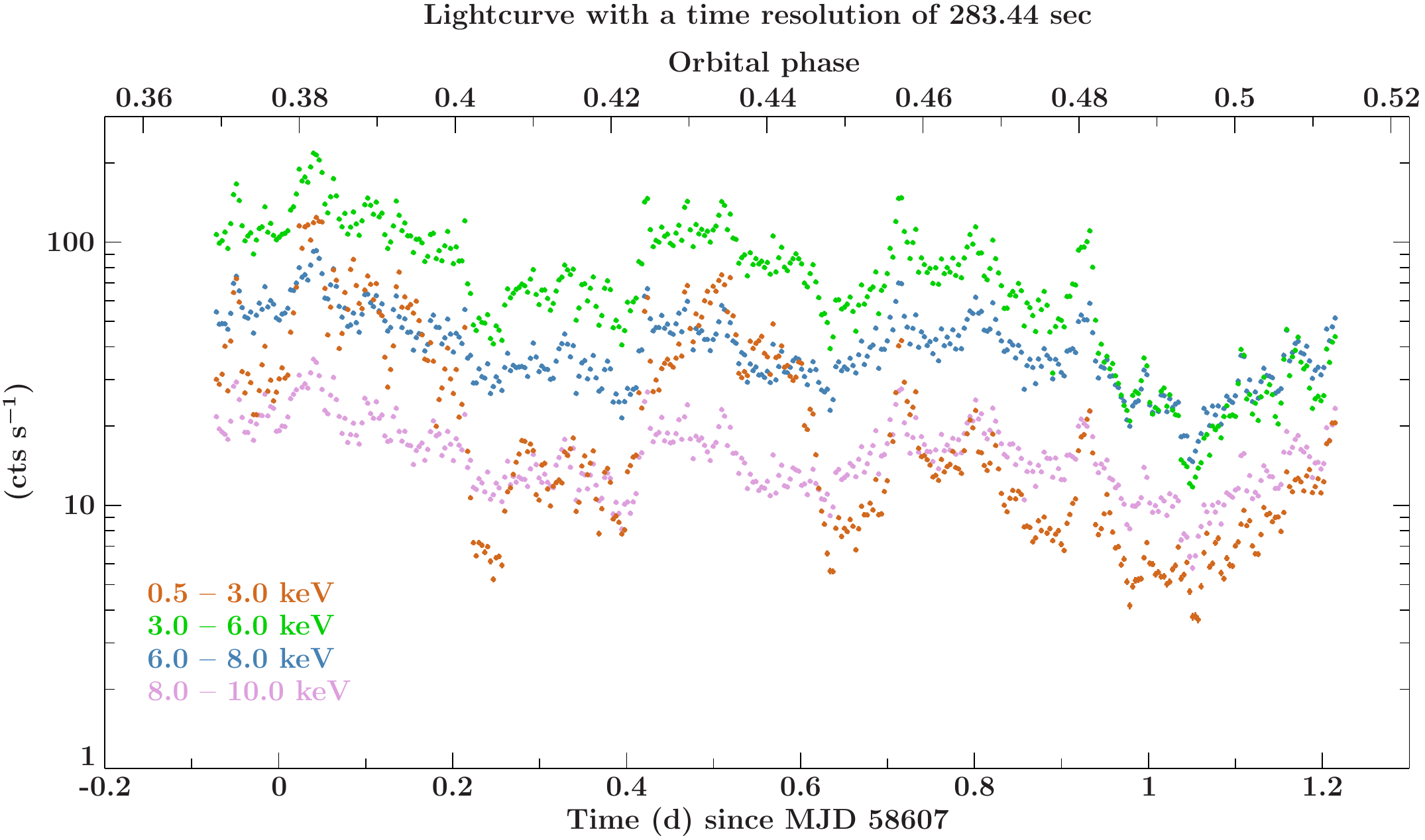}}
    \caption{Light curves for \xmmnewton EPIC-pn in different energy bands with a time resolution of $P=283.44\,\mathrm{s}$. The count rate is plotted against the orbital phase and the time of the observation, y-axis in logarithmic scale. The energy bands are 0.5--3.0, 3.0--6.0, 6.0--8.0, 8.0--10.0\,keV.}
    \label{fig:xmm_lc}
\end{figure*}

We present the \xmmnewton EPIC-pn light curves in the above mentioned energy bands in Fig.~\ref{fig:xmm_lc} . All light curves show the same features and variability in all energy bands, but the flares and low states are more prominent at low energies particularly in the 0.5--3.0\,keV energy band where the brightest flare is approximately five times higher than the average count rate in that band ($\sim$24\,counts$\,$s$^{-1}$). We can notice a corresponding trend for the low state at $T_{\mathrm{obs}} \approx 58607.24 \ \mathrm{MJD}$ being roughly five times smaller than the average count rate in the 0.5--3.0\,keV band. The long and broad flare is visible in all energy bands but again is much more prominent in the 0.5--3.0\,keV energy band and shows a rather stable plateau at $\sim$40\,counts$\,$s$^{-1}$. In the two highest energy bands, the overall count rate stays rather stable around the mean value of each individual light curve. This phenomenon has been observed in a similar energy band, 1.0--3.0\,keV, in \citet{Martinez_2014} and has been associated to a stable spectral shape from the unabsorbed source.

\subsection{Hardness ratios}\label{sec:hardness}
\begin{figure}
    \centering
    \centerline{\includegraphics[trim=0cm 0cm 0cm 0.5cm, clip=true, width=1.0\linewidth]{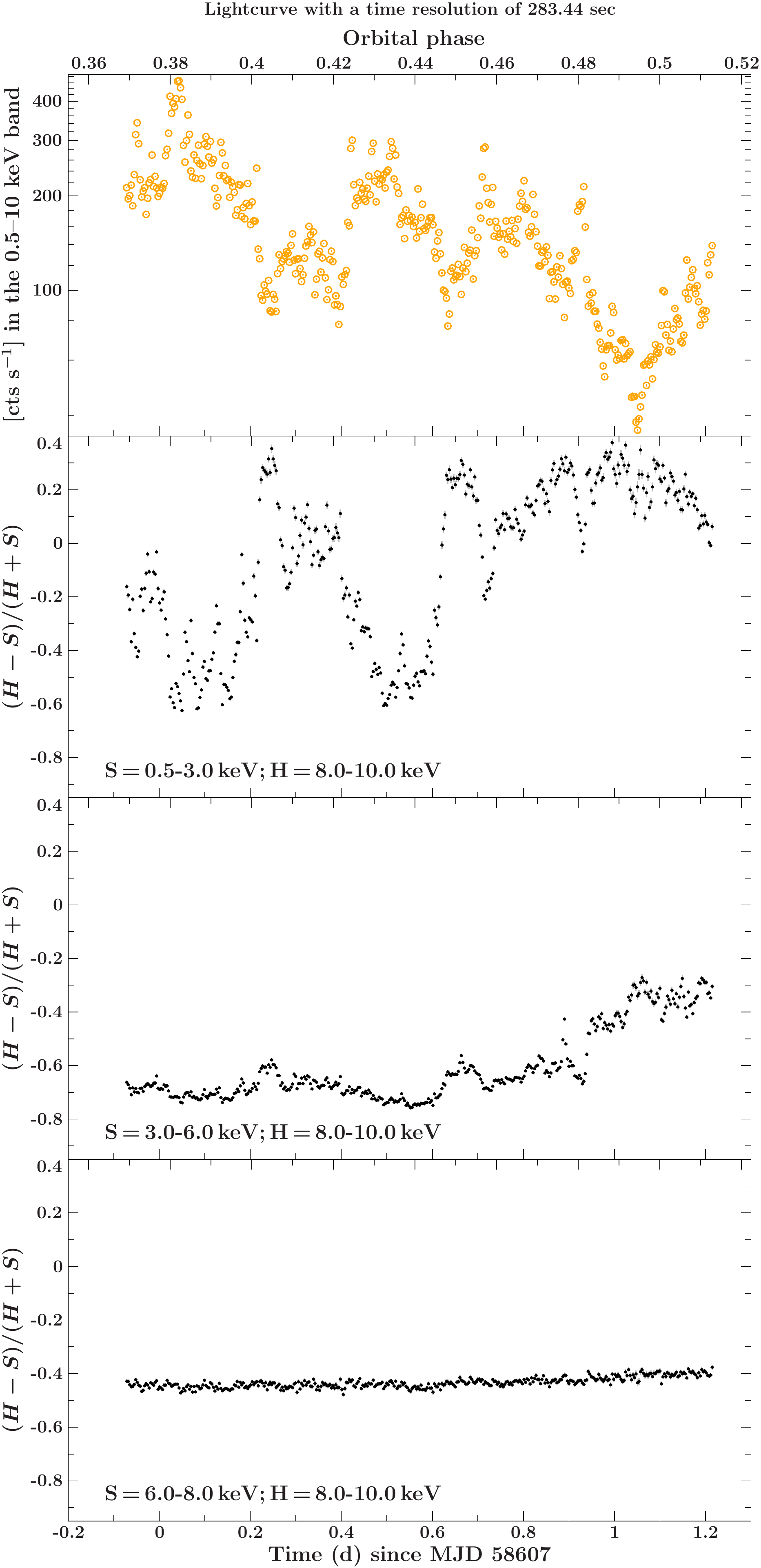}}
    \caption{Light curve and hardness ratios for \xmmnewton EPIC-pn. First panel shows the overall count rate in the 0.5--10\,keV energy band as in Fig.~\ref{fig:both_lc}. Following panels show the hardness ratios between the mentioned energy bands. The time resolution is $P=283.44$ s.}
    \label{fig:hardness_ratios}
\end{figure}

For a more quantitative study of the source variability and in particular in an attempt to determine the origin of the variability shown in Vela X-1, we present in Fig.~\ref{fig:hardness_ratios} the \xmmnewton EPIC-pn hardness ratios. 

In the second panel of Fig.~\ref{fig:hardness_ratios}, we can observe a mirrored version of the light curve for the hardness ratio between the $S$ = 0.5--3.0\,keV and $H$ = 8.0--10.0\,keV bands. The minima of the hardness ratio correspond to the maxima of the light curve and vice-versa. This shows that the observed flares of the light curve happen during the softening of the spectral shape, thus a contribution from low-energy photons mainly as seen in Fig.~\ref{fig:xmm_lc}. On the opposite, low states corresponds to a hardening of the spectral shape so a contribution from high-energy photons mainly. In the third panel, the hardness ratio between the $S$ = 3.0--6.0\,keV and $H$ = 8.0--10.0\,keV bands stays rather constant until $T_{\mathrm{obs}} \approx 58607.60 \ \mathrm{MJD}$ ($\phi_{\mathrm{orb}} \approx 0.44$), with a local maximum at $T_{\mathrm{obs}} \approx 58607.24 \ \mathrm{MJD}$ ($\phi_{\mathrm{orb}} \approx 0.40$). The overall hardness ratio starts to increase after $T_{\mathrm{obs}} \approx 58607.60 \ \mathrm{MJD}$ ($\phi_{\mathrm{orb}} \approx 0.44$) until $T_{\mathrm{obs}} \approx 58607.93 \ \mathrm{MJD}$ ($\phi_{\mathrm{orb}} \approx 0.48$) followed by a steeper increasing slope until the end of the observation. This was expected from \citet{Diez_2022} as we analysed similar energy bands for the hardness ratio: the 3.0--5.0\,keV and 20.0--30.0\,keV bands which also correspond to the low-energy and continuum parts of the spectrum respectively. Finally, in the last panel of Fig.~\ref{fig:hardness_ratios}, the hardness ratio between the 6.0--8.0\,keV and 8.0--10.0\,keV bands is constant throughout the whole observation, showing no variation in the continuum as in \citet{Diez_2022} and \citet{Martinez_2014}. 

To summarise, it seems that the spectral shape gradually changes at low energies. It is particularly visible between 0.5\,keV and 3.0\,keV. This could be associated to changes of behaviour of the absorbing material while the continuum emission from the neutron star seems to be stable. 

\begin{figure}
    \centering
    \centerline{\includegraphics[trim=0cm 0cm 0cm 1.1cm, clip=true, width=1.0\linewidth]{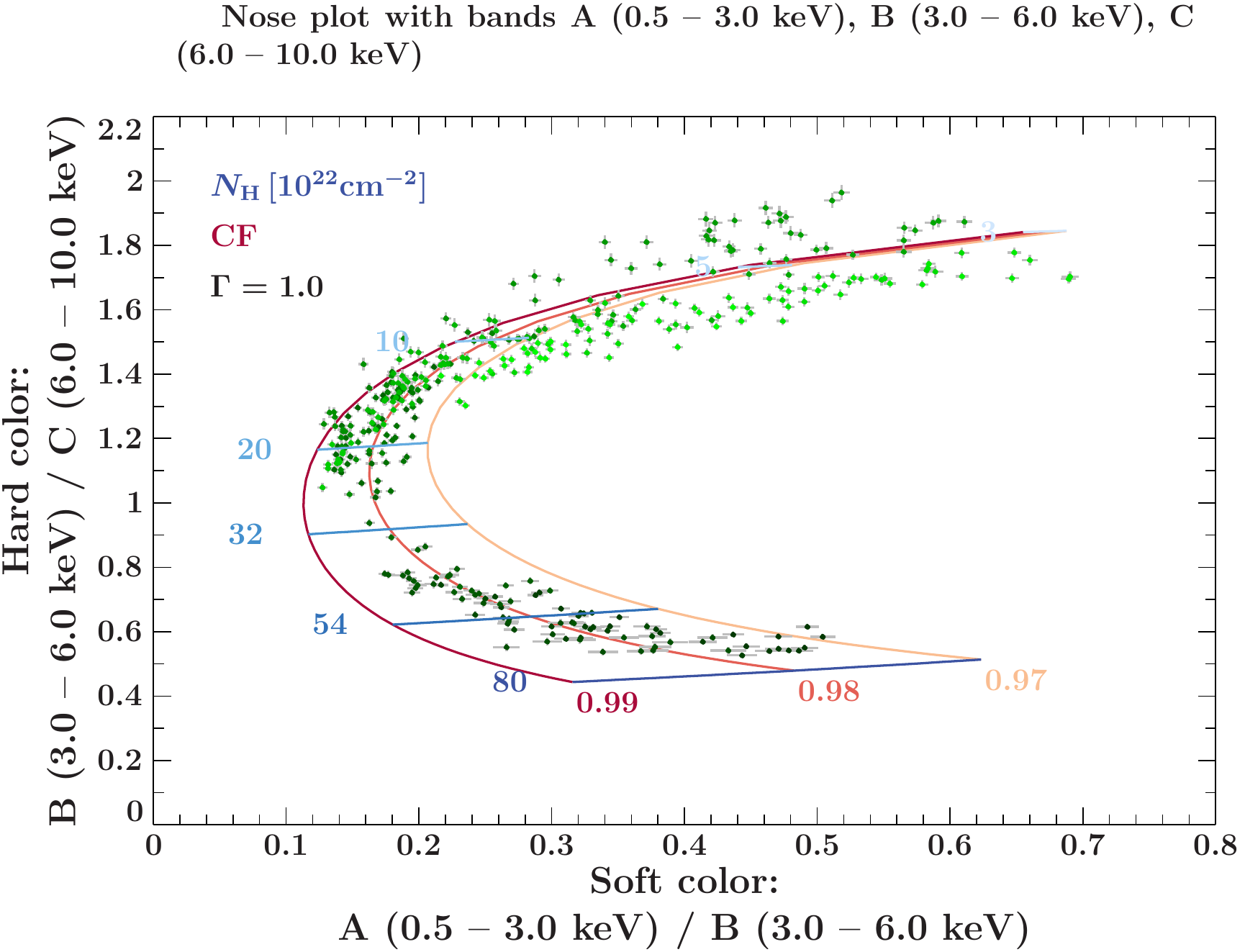}}
    \caption{\xmmnewton EPIC-pn colour-colour diagram. The data points represent the ratios of the light curves in hard colour depending on soft colour, from beginning (light green) to end (dark green) of the observation.
    We also show the theoretical expectation for different covering fractions $\mathrm{CF}$ (shades of red, varying from 0.97 to 0.99) and absorption column densities $N_{\mathrm{H,1}}$ (shades of blue, varying from $3 \times 10^{22}\,\mathrm{cm}^{-2}$ to $80 \times 10^{22}\,\mathrm{cm}^{-2}$) using our partial covering model from Eq.~\ref{eq:final_model}. We used a photon index $\Gamma$ of 1.0. More details about the simulation and its interpretation are given in Sect.~\ref{section:discuss_orbital_evol}.}
    \label{fig:nose_plot_0.5-3.0_3.0-6.0_6.0-10.0}
\end{figure}

A further hint towards the role of the wind absorption comes from a colour-colour diagram (Fig.~\ref{fig:nose_plot_0.5-3.0_3.0-6.0_6.0-10.0}) that shows a typical 'nose'-like shape that has been previously associated with variable absorption in stellar wind especially in Cyg~X-1 \citep{Nowak_2011a,Hirsch_2019,Grinberg_2020,Lai_2022}. We further discuss our modelling of the colour-colour diagram and its implications for the wind properties in Sect.~\ref{section:discuss_orbital_evol}.

To explore the source behaviour further, a spectral analysis of the absorption column density on shorter timescales is necessary. 

\section{Spectral analysis}
\label{section:spectrum_analysis}

\subsection{Spectral model definition}

\subsubsection{Continuum shape}

Since this current \xmmnewton EPIC-pn work follows results from simultaneous observation analysed in our previous NuSTAR work, we decide to use the same continuum model for direct comparisons and homogeneous continuity of the work. The final continuum model we obtained is the following and we refer to \citep{Diez_2022} for an in-detail discussion on how this model was obtained:

\begin{align}
\label{eq:final_model_nustar}
\begin{split}
I(E) = & N_{\rm{H,2}} \times (\mathrm{CF} \times N_{\rm{H,1}} + (1-\mathrm{CF})) \\
& \times (F(E) \times \rm{CRSF,F} \times \rm{CRSF,H} + \rm{FeK\alpha} + 10\,\rm{keV}).
\end{split}
\end{align}

The parameter $N_{\rm{H,2}}$ accounts for the absorption column density from the interstellar medium and is fixed to $3.71 \times 10^{21}\,\rm{cm^{-2}}$ using NASA's HEASARC $\rm{N_H}$ tool website\footnote{\url{https://heasarc.gsfc.nasa.gov/cgi-bin/Tools/w3nh/w3nh.pl}} \citep{HI4PI_2016}. The absorption $N_{\rm{H,1}}$ corresponds to the stellar wind embedding the neutron star and is a free parameter. Those two absorption components are described by the \texttt{tbabs} model \citep{Wilms_2000} with the corresponding abundances and cross-sections from \citet{Verner_1996}.

The covering fraction $\mathrm{CF}$ quantifies the clumpy structure of the stellar wind and ranges between 0 (the obscurer does not cover the source) and 1 (the obscurer is fully covering the source). This is characteristic of a partial covering model, of which several flavours can be found in the literature \citep[e.g.][]{Martinez_2014,Fuerst_2014b,Malacaria_2016a}, and had been found to be the best description of the clumpy absorber in Vela X-1. 

The function $F(E)$ describes the spectral continuum of the accreting neutron star. It is empirically described by a power law with a high-energy cutoff \citep[see e.g. ][]{Staubert_2019}. Several models from the literature may account for the description of the high-energy cutoff. Our best results were obtained with the \texttt{FDcut} high-energy cutoff so that:

\begin{align}
\label{eq:fdcut}
F(E) \propto E^{-\Gamma} \times \left(1+ \exp \left(\frac{E-E_{\rm{cut}}}{E_{\rm{fold}}} \right) \right) ^{-1},
\end{align}
where $\Gamma$, $E_{\rm{cut}}$ and $E_{\rm{fold}}$ stand for the photon index, the cutoff energy and the folding energy respectively.

In the spectrum of Vela X-1, two Cyclotron Resonant Scattering Features (CRSFs, or cyclotron lines) are present with a prominent harmonic line at $\sim$55\,keV and a weaker fundamental at $\sim$25\,keV \citep{Kendziorra_1992,Kretschmar_1997,Orlandini_1998,Kreykenbohm_1999a,Kreykenbohm_2002, Fuerst_2014a, Diez_2022}. Those features are typical of highly magnetised neutron stars and can be observed in the source X-ray spectrum as broad absorption lines. CRSFs result from resonant scattering of photons by electrons in strong magnetic fields from the ground level to higher excited Landau levels followed by radiative decay \citep[see][for a review]{Staubert_2019}. We described the CRSFs using two multiplicative Gaussian absorption lines with the \texttt{gabs} parameter in XSPEC and corresponding to the fundamental and harmonic CRSF so that:  
\begin{align}
    \rm{CRSF}(E) = \exp\left[-\left(\frac{d}{\sigma\sqrt{2\pi}}\right)\exp\left(-0.5\left(\frac{E-E_{cyc}}{\sigma}\right)^2\right)\right],
\end{align}
where $d$ is the line depth and $\sigma$ the line width. 

The fluorescent emission line associated with $\mathrm{FeK\alpha}$ was modelled with a narrow Gaussian line component with the \texttt{egauss} parameter in XSPEC around 6.4\,keV with width and flux left to vary. 

The 10 keV feature is described by a broad Gaussian line component in absorption. The physical origin of this feature is still unknown but we discussed in \citet{Diez_2022} the presence of this feature in Vela X-1. 

\subsubsection{Line emission}

Thanks to the low-energy coverage permitted by \xmmnewton EPIC-pn and a better resolution than \nustar between 3 and 10\,keV, we have now access to new features that we need to include in our model.

In our previous analysis, the fluorescent emission line associated with the $\mathrm{FeK\beta}$ could not be resolved from the $\mathrm{FeK\alpha}$ emission line. This is now possible with the \xmmnewton EPIC-pn energy resolution and we model it with a Gaussian component around 7.1\,keV with flux left free to vary but fixed width.

Especially in the most absorbed spectra of our observation, we can now also identify multiple emission lines between 0.5\,keV and 4\,keV (see last panel of Fig~\ref{fig:evol_spectra}). The absorption from the stellar wind is very strong towards the end of the observation (i.e. towards late orbital phases, see \citealt{Diez_2022}), hence the strong continuum emitted by the neutron star is heavily absorbed and reveals the emission lines normally subsumed in the continuum when the absorption is less strong. 

\begin{figure}
    \centering
    \centerline{\includegraphics[trim=0cm 0cm 0cm 0cm, clip=true, width=1.\linewidth]{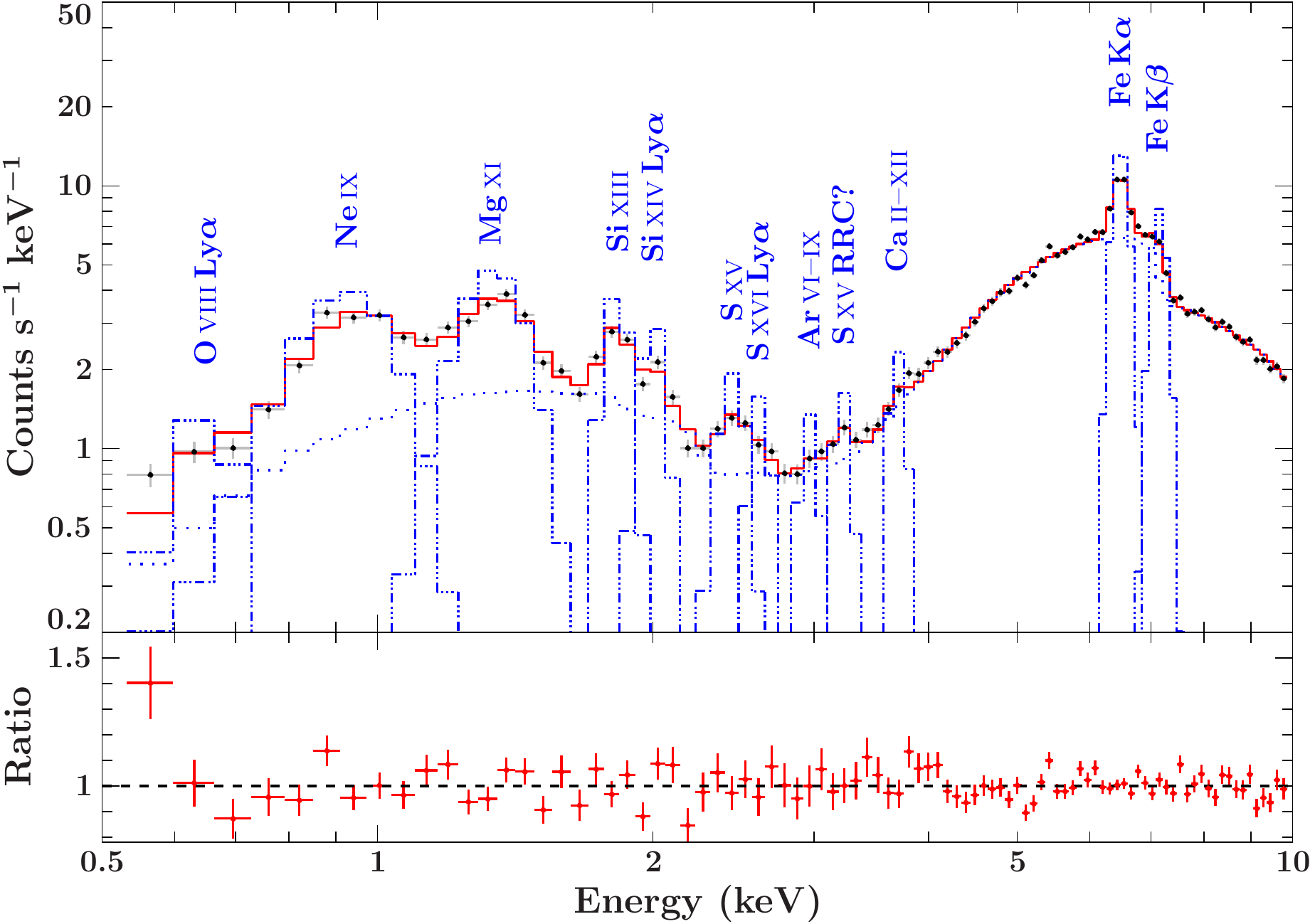}}
    \caption{Example \xmmnewton EPIC-pn spectrum (black datapoints): we show the last and most absorbed \nustar-orbit of our observation. We indicate the individual model components including all lines detected in this dataset (blue dot-dashed Gaussians) and the absorbed continuum (blue dotted line). We refer to Sect.~\ref{subsection:spectral_analysis_orbit_XMM_NuSTAR} for a detailed description of the model and to Table~\ref{tab:emission_line_details} for the soft lines.}
    \label{fig:plot_comps}
\end{figure}

To help us identifying the energy of individual observed features, we based our search on \chandra/HETGS results of \citet{Amato_2021a}, who analysed Vela X-1 at orbital phase $\phi_{\mathrm{orb}} \approx 0.75$, which is even more affected by the stellar wind (see Fig.\ref{fig:vela_sketch}). Because of the EPIC-pn limited energy resolution, Doppler shifts with orbital phase and triplets or faint lines cannot be resolved in this work. To search for the lines in EPIC-pn spectra, we had to fix the features (such as the CRSFs and the 10 keV feature) and continuum parameters (such as $E_{\rm{cut}}$ and $E_{\rm{fold}}$) that are not covered by the EPIC-pn instrument. We fixed them to the average \nustar values from \citet{Diez_2022} for highly absorbed spectra to have an accurate description of the continuum to focus on the lines description. We performed this search 'by hand' fitting Gaussian components to observed line features in the most absorbed time-resolved spectra of the observation until we obtain a satisfactory reduced chi-square. An example of such a spectrum is shown in Fig.\ref{fig:plot_comps}. The energies of the narrowest lines (\NeX\,\Lya, \MgXII\, \SiXIV\,\Lya, \SXVI\,\Lya, \CaIIXII\,K$\alpha$) have to be fixed to previous studies and their widths to $10^{-6}$\,keV.

A list of the soft lines identified in this work and a comparison with previous studies is shown in Table~\ref{tab:emission_line_details}. It has to be considered that this work does not aim to do an in-depth study of the emission lines variability but is oriented to an absorption study of the stellar wind in Vela X-1. The instrumental issues described in Sect.~\ref{section:data_reduc} with the limited energy resolution of EPIC-pn can explain the discrepancies obtained for some lines in comparison with previous work, in particular for \OVIII\,\Lya\, and \CaIIXII\,K$\alpha$. The \ArVIIX\, and \CaIIXII \, $\mathrm{K\alpha}$ lines were not detected in \citet{Amato_2021a}. However, those lines were identified in \citet{Schulz_2002}, \citet{Goldstein_2004} and \citet{Watanabe_2006} for Vela X-1 and in \citet{Fuerst_2011} for the HMXB GX 301--2. The Ca and Ar different charge states cannot be resolved so we give in Table~\ref{tab:emission_line_details} the potential candidates as in \citet{Schulz_2002} and \citet{Watanabe_2006}. 

\begin{table*}[]
\renewcommand{\arraystretch}{1.1}
\caption{Details of soft emission lines between 0.5\,keV and 4\,keV.}
\label{tab:emission_line_details}
\begin{center}
\begin{small}
\begin{tabular}{llll}    
\hline\hline
Line &
Detected energy &
Reference energy &
\multicolumn{1}{l}{Identified energy}\\
& 
from previous work (keV) &
(keV)
& 
\multicolumn{1}{l}{for this work (keV)}\\
\hline
\OVIII \, \Lya  & $0.6538^{+0.0005}_{-0.0011}{}^{\,(a)}$  & $0.6541^{\,(c)}$ & $0.6211^{+0.0130}_{-0.0011}$\\
\NeIX \, (f, i, r)  & $0.90460\pm{0.00033}$/$0.91454\pm{0.00034}$/$0.92154\pm{0.00034}{}^{\,(b)}$ & $0.905/0.915/0.922{}^{\,(c)}$ & $0.928^{+0.011}_{-0.012}$\\
\NeX \, \Lya   & $1.02130^{+0.00015}_{-0.00014}{}^{\,(b)}$ & $1.02196{}^{\,(d)}$ & 1.02130 (fixed)\\
\MgXI \, (f, i, r)  & $1.3305\pm{0.0002}$/$1.3426^{+0.0003}_{-0.0002}$/$1.3517^{+0.0002}_{-0.0003}{}^{\,(b)}$ & $1.3311/1.3431/1.3522{}^{\,(c)}$ & $1.338^{+0.012}_{-0.019}$\\
\MgXII \, \Lya  & $1.4720\pm{0.0002}{}^{\,(b)}$ & $1.4723{}^{\,(d)}$ & 1.4720 (fixed) \\
\SiXIII \, (f, i, r)  & $1.8388\pm{0.0002}$/$1.8536\pm{0.0002}$/$1.8643\pm{0.0002}{}^{\,(b)}$  & $1.8382/1.8530/1.8648{}^{\,(e)}$ & $1.823^{+0.014}_{-0.013}$\\
\SiXIV \, \Lya  & $2.0049\pm{0.0003}{}^{\,(b)}$ & $2.0056{}^{\,(f)}$ & 2.0049 (fixed) \\
\SXV \, (f, i, r)  & $2.4287^{+0.0007}_{-0.0008}$/$2.4463^{+0.0007}_{-0.0009}$/$2.4590^{+0.0006}_{-0.0009}{}^{\,(b)}$ & $2.4291/2.4463/2.4606{}^{\,(e)}$ & $2.439^{+0.029}_{-0.027}$ \\
\SXVI \, \Lya  & $2.6207^{+0.0016}_{-0.0017}{}^{\,(b)}$ & $2.6196{}^{\,(f)}$ & 2.6207 (fixed) \\
\ArVIIX \, & $2.9661^{+0.0043}_{-0.0099}{}^{\,(g)}$ & $2.9619-2.9675{}^{\,(h)}$ & 2.9661 (fixed) \\ 
\SXV\,RRC?  &  & $3.224{}^{\,(c)}$ & $3.23^{+0.04}_{-0.06}$ \\
\CaIIXII \,  $\mathrm{K\alpha}$  & $3.6905^{+0.0022}_{-0.0009}{}^{\,(i)}$ & $3.6911-3.7110{}^{\,(h)}$ & $3.822^{+0.019}_{-0.102}$\\
\hline
\multicolumn{4}{l}{(f, i, r): Referring to forbidden, intercombination and resonance lines respectively.}\\
\multicolumn{4}{l}{(fixed): to previous detected energies.}\\
\multicolumn{4}{p{0.9\linewidth}}{Note: To convert from \angstrom\, in previous works to keV, we compute $E {\mathrm{[keV]}} = hc \div \lambda{\mathrm{[\angstrom]}}$ where $hc=12.39842$ (with values for $h$, $c$ and $e$ from CODATA 2018, \citealt{CODATA_2018}) and round to relevant significant digits.}\\
\multicolumn{4}{p{0.9\linewidth}}{$^{\,(a)}$ \citet{Lomaeva_2020a}, $^{\,(b)}$ \citet{Amato_2021a}, $^{\,(c)}$ \citet{Drake_1988} , $^{\,(d)}$ \citet{Erickson_1977}, $^{\,(e)}$ \citet{Hell_2016}, $^{\,(f)}$ \citet{GarciaMack_1965}, $^{\,(g)}$ \citet{Schulz_2002}, $^{\,(h)}$ \citet{House_1969}, $^{\,(i)}$ \citet{Watanabe_2006} }\\

\end{tabular}
\end{small}
\end{center}
\renewcommand{\arraystretch}{1.1}
\end{table*}

We also detect a 3.2\,keV line that has not been reported in previous work. This feature can most likely be attributed to the \SXV\,RRC, with its ionisation potential of 3.224\,keV \citep{Drake_1988}. While lower charge states of Ar are present (2.9661\,keV), the line energies for both He-like \ArXVII\,r at 3.140\,keV \citep{Drake_1988} and H-like \ArXVIII\,\Lya\,at 3.321\,keV \citep{GarciaMack_1965} are too far away to reasonably match the fitted line energy. The same is true for the H-like \SXVI\,\Lyb\,and \Lyg\,lines at 3.106  and 3.276\,keV \citep{GarciaMack_1965}, respectively, even though the \SXVI\,\Lya\,feature is clearly detected. If this feature is indeed caused by the \SXV\,RRC, we would expect RRC features from the more abundant ions as well. However, at the low resolution of the CCD spectra, many of the other RRC candidates are too blended to allow for a clear detection; for instance, the \SiXIII\,RRC at 2.438\,keV \citep{Drake_1988} blends with the \SXV\,He$\alpha$ complex, the \NeX\,RRC at 1.362\,keV \citep{GarciaMack_1965} blends \MgXI\,\Hea, and the \OVIII\,RRC at 0.871\,keV \citep{GarciaMack_1965} with the \NeIX\,\Hea\,lines \citep[see e.g.][]{Sako_1999a}.

Finally, our final and best-fit model that we use for the time-resolved spectroscopy in this work is:

\begin{align}
\label{eq:final_model}
\begin{split}
I(E) =  &N_{\rm{H,2}} \times [(\mathrm{CF} \times N_{\mathrm{H,1}} + (1-\mathrm{CF})) \times (F(E) \times \rm{CRSF,F} \\
&\times \rm{CRSF,H}) 
 + \rm{FeK\alpha} + \rm{FeK\beta} + 10\,\rm{keV} + \mathrm{Soft \ lines}].
 \end{split}
\end{align}

\subsection{\nustar-orbit-by-orbit analysis}
\label{subsection:spectral_analysis_orbit_XMM_NuSTAR}

We can now perform the analysis on shorter timescales with both \xmmnewton EPIC-pn and \nustar to access the variability of the stellar wind at low energies. Combining both datasets, and thus increasing the covered spectral range, limits the impact of possible degeneracy between the power law slope and absorption strength. \nustar data are especially crucial at high absorption, when it is especially difficult to constrain the continuum with \xmmnewton only.

\begin{figure*}
    \centering
    \centerline{\includegraphics[trim=0cm 0cm 0cm 0cm, clip=true, width=0.6\linewidth]{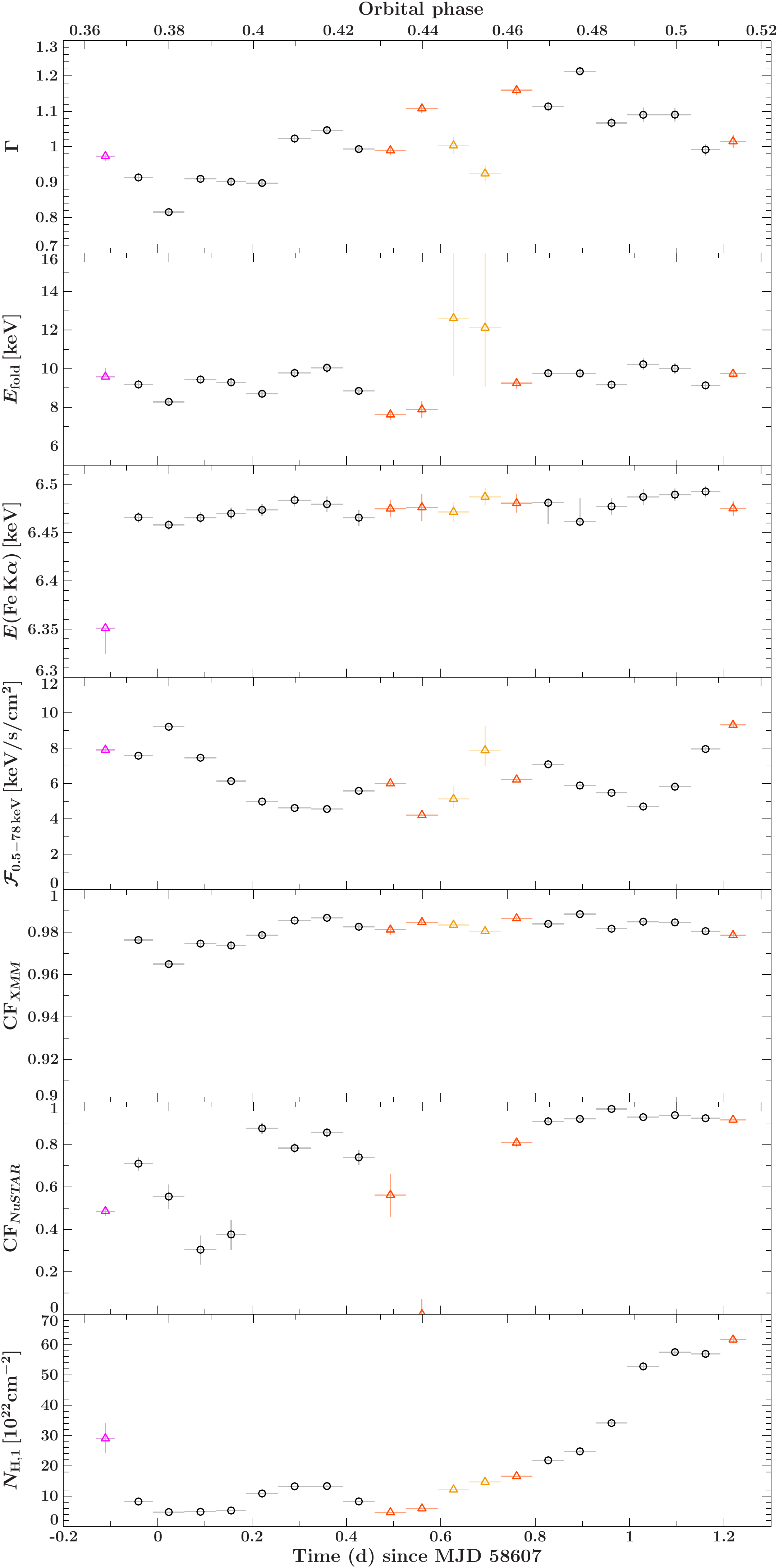}}
    \caption{Results of the \nustar-orbit-by-orbit analysis with \nustar and \xmmnewton as functions of time, showing also
    the corresponding binary orbital phase. From top to bottom: photon index ($\Gamma$), folding energy ($E_{\mathrm{fold}}$) in keV, energy of the $\mathrm{FeK\alpha}$ line in keV, unabsorbed flux $\mathcal{F}_{\rm{0.5-78 \ keV}}$ in $\rm{keV \, s^{-1} \, cm^{-2}}$, covering fractions with \nustar ($\mathrm{CF_{\nustar}}$) and \xmmnewton EPIC-pn ($\mathrm{CF_{\textsl{XMM}}}$), absorption from the stellar wind $N_{\rm{H,1}}$ in $10^{22}\,\rm{cm^{-2}}$. Circles: data fully covered by both \xmmnewton EPIC-pn and \nustar. Triangles: data missing coverage from one of the two instruments. In magenta: data only covered by \nustar (\nustar-orbit 1); in orange: data fully covered by one instrument but partially by the second one (\nustar-orbits 10, 11, 14 and 21); in yellow: data only covered by \xmmnewton EPIC-pn (\nustar-orbits 12 and 13).}
    \label{fig:some_param_vs_time}
\end{figure*}

In \citet{Diez_2022}, we extracted a spectrum for every orbit of \nustar around the Earth, called '\nustar-orbit' for the remainder of the paper. This should not be confused with the duration of a binary orbit of the neutron star around its companion. For the simultaneous \xmmnewton EPIC-pn data, we decided to extract a spectrum using the same Good Time Intervals (GTIs) used for the \nustar-orbit-by-orbit analysis with \nustar data in order to have a broad X-ray band description of the stellar wind on the same timescale. Due to different start and stop time of the observations, there are \nustar-orbits without simultaneous \xmmnewton EPIC-pn data (see Fig.~\ref{fig:both_lc}). Moreover, during the \nustar campaign, data were lost because of ground station issues, resulting in no or limited \nustar coverage for parts of our \xmmnewton observation. For the 'missing' \nustar-orbits, we took the average duration of a \nustar-orbit, which lasts $\sim$0.067 MJD ($\sim$5.8\,ks), in order to extract \xmmnewton EPIC-pn spectra. Overall, we have 1 \nustar-orbit covered by \nustar only, 2 \nustar-orbits covered by \xmmnewton EPIC-pn only, 4 \nustar-orbits partially covered by one of the two instruments and 14 \nustar-orbits fully covered by both instruments (Fig.~\ref{fig:some_param_vs_time}). We fit the data using the model from Eq.~\ref{eq:final_model}, with adaptations as necessitated by the different instrumental coverage as discussed in the following.

Firstly, for the \nustar-orbits covered by both \nustar and \xmmnewton EPIC-pn, we use a floating cross-normalisation parameter, $\mathcal{C}_{\mathrm{\nustar}}$, in order to give the relative normalisation between the two \nustar detectors FPMA and FPMB with the \xmmnewton EPIC-pn instrument. The difference between FPMA and FPMB is of the order of $\sim$2\% so we can safely assume one normalisation constant to account for both focal plane modules for simplicity. 

As discussed in Sect.~\ref{section:data_reduc}, cross-calibration issues between \nustar and \xmmnewton EPIC-pn are impacting the analysis. To correct the observed up-turn in the \nustar data at $\sim$3\,keV, we applied two different covering fractions $\mathrm{CF_{\textsl{XMM}}}$ and $\mathrm{CF_{\nustar}}$ for \xmmnewton EPIC-pn and \nustar respectively. We do not fix $\mathrm{CF_{\nustar}}$ to previous values from \citet{Diez_2022} as this current work benefits from low-energy coverage with \xmmnewton which gives us better constraints on $N_\mathrm{H,1}$, therefore on $\mathrm{CF_{\nustar}}$ since those two parameters were found to be strongly correlated in Fig.~7 of \citet{Diez_2022}. To correct the shift in the $\mathrm{FeK\alpha}$ emission line, we apply a gainshift to the \nustar data in order to align on the iron line energy found with \xmmnewton EPIC-pn. 

We also introduce another cross-normalisation constant $\mathcal{C}_{\mathrm{Fe}}$ to account for the flux difference observed in \nustar relatively to \xmmnewton EPIC-pn in the emission lines that are covered by both observatories: $\mathrm{FeK\alpha}$ and $\mathrm{FeK\beta}$.  

We fix the soft emission lines energy and width to the values estimated when analysing the \xmmnewton EPIC-pn spectrum at high absorption (see Table~\ref{tab:emission_line_details}) as they are not expected to change significantly with time even if the source is highly variable \citep{Grinberg_2017}. The fluxes of the emission lines are left free as their prominence changes depending on the local absorption. 

We fix the CRSFs parameters and the energy of the 10 keV feature to the values of their corresponding \nustar-orbit we obtained in our previous analysis in \citet{Diez_2022} to help constraining the low-energy part of the continuum. Degeneracies between $E_{\mathrm{CRSF,F}}$ and $E_{\rm{cut}}$ are expected due to their proximity as seen in \citet{Diez_2022} thus we also fix the cutoff energy in the same way. 

Secondly, for the two \nustar-orbits only covered by \xmmnewton EPIC-pn (\nustar-orbits number 12 and 13), we fix the CRSFs parameters, the energy of the 10 keV feature $E_{\mathrm{10\,keV}}$ and the cutoff energy $E_{\rm{cut}}$ to the closest \nustar-orbit values. These parameters cannot be ignored during the fitting because they modify the shape of the broadband continuum, thus fixing them to the values of the closest \nustar-orbit is the most accurate and meaningful solution. We do not use $\mathcal{C}_{\mathrm{\nustar}}$, gainshift, $\mathrm{CF_{\nustar}}$, nor $\mathcal{C}_{\mathrm{Fe}}$ as there is no need to correct for cross-calibration because no coverage from \nustar is available for those \nustar-orbits.

Thirdly, for the \nustar-orbit that is covered by \nustar only (\nustar-orbit number 1), the multiple soft emission lines and the $\mathrm{FeK\beta}$ line identified with \xmmnewton EPIC-pn are not resolved by \nustar alone. However, contrary to the CRSFs, the 10 keV feature and cutoff energy previously, those emission lines do not impact the shape of the overall continuum of Vela X-1 therefore they can be safely ignored for fitting the data for simplicity. We tried to fix them to the values of the closest following \nustar-orbit but no significant difference when comparing the residuals could be highlighted.

We present the results of the \nustar-orbit-by-orbit analysis in Fig.~\ref{fig:some_param_vs_time} focusing on the parameters of interest. An example of a fitted spectrum extracted during one \nustar-orbit with both \xmmnewton EPIC-pn and \nustar is given in Fig.~\ref{fig:ex_spec_XMM_NuSTAR}. However, we caution that the different instrumental coverage of individual spectra may result in artificial parameter behaviour, for instance, outliers.

Discrepancies between \nustar and \xmmnewton EPIC-pn are visible in the cross-normalisation parameter $\mathcal{C_{\mathrm{\nustar}}}$ reaching $\sim$20\% (not considering outliers) essentially explained by the low-energy up-turn in \nustar (see Fig.~\ref{fig:ex_spec_XMM_NuSTAR}). Moreover, the average energy gainshift of \nustar relatively to \xmmnewton EPIC-pn is of -87\,eV. This particularly impacts the iron line region as it is covered by both instruments. Therefore, the energy of the iron line shown in the third panel of Fig.~\ref{fig:some_param_vs_time} is given with respect to the \xmmnewton EPIC-pn values, the gainshift has to be added to retrieve the \nustar values, with the exception of the magenta triangle outlier of \nustar-orbit 1 which is only covered by \nustar.

Significant variability can be observed in the presented parameters. In particular, $N_\mathrm{H,1}$ increases by a factor of 6 between the beginning and the end of the observation showing a very clear rise of the stellar wind absorption at $\phi_{\mathrm{orb}} \approx 0.44-0.49$. The energy of the iron line stays rather stable around $\sim$6.48\,keV but shows local minima which seem to be anti-correlated with the flux in the 3--10\,keV energy band. This anti-correlation with flux appears to be similar for the photon index showing local dips during flares. The photon index varies overall between 0.8 and 1.2, the change of spectral shape being associated to changes of absorption density in the stellar wind and possible degeneracy with the amount of absorption. Further investigation and details will be discussed in Sect.~\ref{section:discussion}.

While the covering fraction with \xmmnewton EPIC-pn $\mathrm{CF_{\textsl{XMM}}}$ remains very stable around 0.98, the covering fraction with \nustar is much more variable ranging between 0.3 and 0.98 (we note the different y-axis range compared to $\mathrm{CF_{\textsl{XMM}}}$) as in \citet{Diez_2022} when analysing \nustar data alone. This is due to the up-turn in \nustar data at low energies (see Fig.~\ref{fig:ex_spec_XMM_NuSTAR}).  We discuss tests performed to assess some possible sources of this effect, in particular a possible contribution from dust scattering, in Sect.~\ref{appendix:cross-instrumental_issues}. We are unable to find a plausible physical explanation and have thus to conclude that the problem is due to some remaining calibration effects. In \citet{Diez_2022}, we have previously discussed the problems encountered with the low-energy effective area correction for FPMA \citep{Madsen_2020} in our observation. While these problems should not affect FPMB, they, together with the high variability of the covering fraction as deducted from the \nustar data alone, imply a reduced reliability of the \nustar data for our observation in this energy range. Moreover, looking at Fig.\ref{fig:nose_plot_0.5-3.0_3.0-6.0_6.0-10.0}, it is obvious that a $\mathrm{CF}$ of less than 0.97 would not describe the data. We therefore decide to focus our discussion on \xmmnewton driven $\mathrm{CF}$ values for the remainder of the paper.

\subsection{Pulse-by-pulse: \xmmnewton only}

The sensitivity of \xmmnewton EPIC-pn allows us to extract a spectrum down to the pulse period of the neutron star (283\,s). We perform a spectral analysis for every pulse of the neutron star to explore further variability on shorter timescales. 

Given the cross-instrumental issues between \xmmnewton and \nustar and the different coverage of the overall observation (due to both LEO of \nustar and the loss of data), this analysis is performed on \xmmnewton data only. While for the \nustar-orbit-by-orbit analysis it was still possible to safely determine the individual contribution of each instrument for each GTI and exclude outliers when needed, such an approach is not feasible for the pulse-by-pulse analysis. The pulse-by-pulse results for this observation with \nustar only are presented in \citet{Diez_2022}.

We use again the model from Eq.~\ref{eq:final_model} setting up the initial parameters for each pulse spectrum from the results of the corresponding \nustar-orbit of the \nustar-orbit-by-orbit spectral analysis. Given the low signal of the individual datasets, we fix all parameters but the photon index $\Gamma$, the covering fraction $\mathrm{CF_{\textsl{XMM}}}$, the absorption column density from the stellar wind $N_{\mathrm{H,1}}$, the energy of the fluorescent $\mathrm{FeK\alpha}$ line and the flux of all the emission lines. 

\begin{figure*}
    \centering
    \centerline{\includegraphics[clip=true, width=0.6\linewidth]{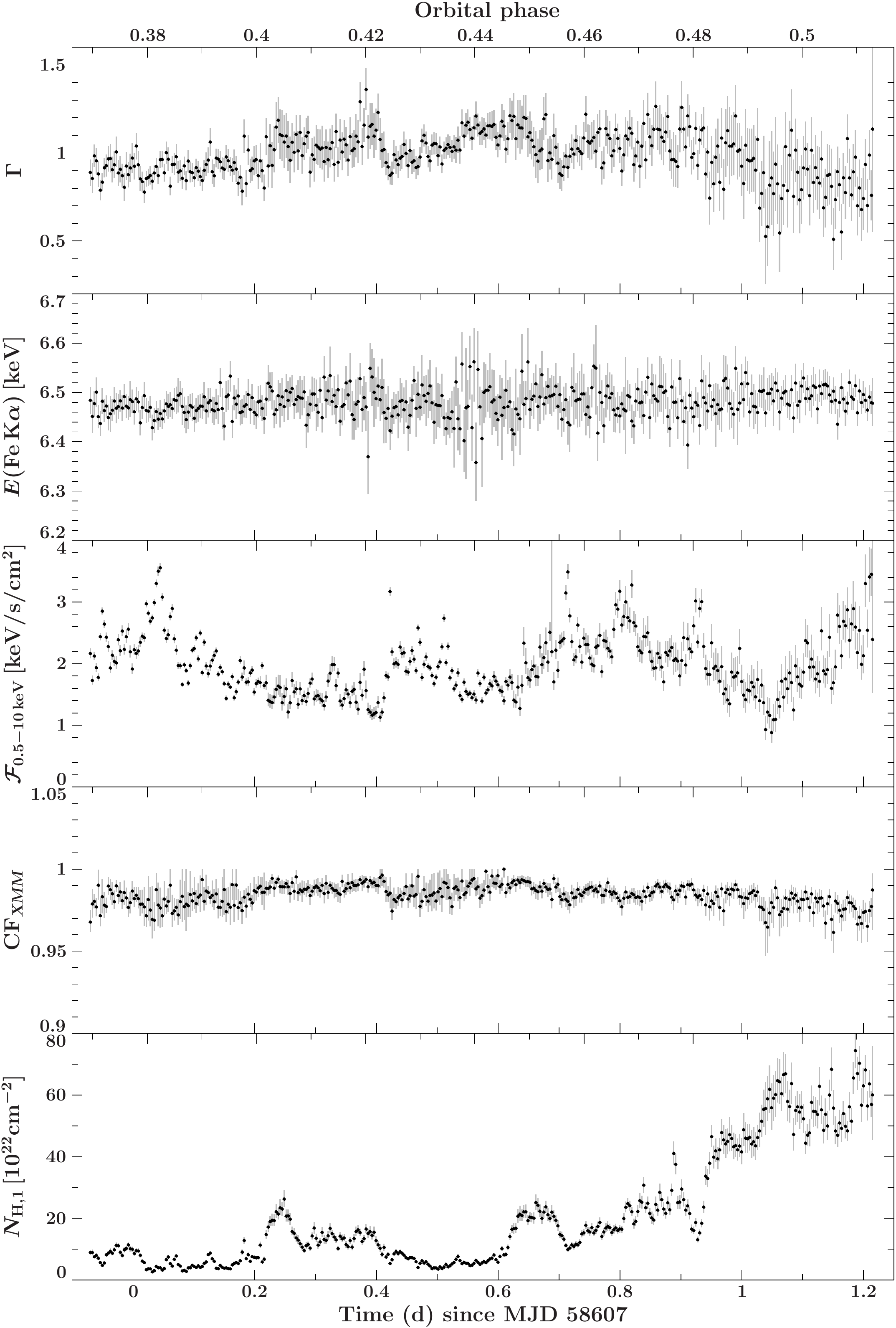}}
    \caption{Results of the pulse-by-pulse analysis as functions of time, showing also the corresponding binary orbital phase. The panels show (from top to bottom): photon index ($\Gamma$), energy of the $\mathrm{FeK\alpha}$ line in keV, unabsorbed flux $\mathcal{F}_{\rm{0.5-10 \ keV}}$ in $\rm{keV \, s^{-1} \, cm^{-2}}$, covering fraction with \xmmnewton EPIC-pn ($\mathrm{CF_{\textsl{XMM}}}$) and absorption from the stellar wind $N_{\rm{H,1}}$ in $10^{22}\,\rm{cm^{-2}}$.}
    \label{fig:some_param_vs_time_pulse}
\end{figure*}

In Fig.~\ref{fig:some_param_vs_time_pulse}, we present the results of the pulse-by-pulse analysis for the \xmmnewton EPIC-pn data. The typical reduced chi-square $\chi^2_{\mathrm{red}}$ of the fittings is $\sim$1.20 and the time resolution of $\sim$283\,s gives us access to much more parameter variability along the orbital phase than in Fig.~\ref{fig:some_param_vs_time}. Again, the rise of the absorption column density $N_{\mathrm{H,1}}$ is clearly visible but with much more local instabilities such as two local episodes of high absorption at orbital phases $\sim$0.40 and $\sim$0.44. The overall track is very similar to the hardness ratio between the 3--6\,keV and 8--10\,keV energy bands on the third panel of Fig.~\ref{fig:hardness_ratios} and even more amplified on the second panel of the same figure for the hardness ratio between the softest and hardest energy bands. This is expected as mainly low-energy photons are absorbed by the stellar wind therefore the underlying spectral shape becomes harder during those high-absorption episodes as explained in Sect.~\ref{section:lightcurves}. On the other hand, episodes of low absorption seem to be associated to flaring periods according to Fig.~\ref{fig:some_param_vs_time_pulse} which could be explained by accretion of clumps on the line of sight of the observer as described in \citet{Diez_2022}. In \citet{Diez_2022}, we observed a correlation between $\mathrm{CF_{\textsl{\nustar}}}$ and $N_{\mathrm{H,1}}$ but also with flux. However, there does not seem to be any correlation between \xmmnewton driven $\mathrm{CF}$ values and $N_{\mathrm{H,1}}$ or flux $\mathcal{F}$ for this work. The photon index $\Gamma$ is quite variable particularly towards the end of the observation. This should be taken with caution as the error bars get larger and the absorption is so strong that some degeneracy between $N_{\mathrm{H,1}}$ and $\Gamma$ may happen.

\section{Discussion: the variable absorber} \label{section:discussion}

\subsection{Evolution of absorption along the binary orbital phase: the onset of wakes}

The main finding of our paper is the detailed analysis of the rise of the $N_\mathrm{H,1}$ value and thus of the absorption in the stellar wind that we interpret as the onset of the wakes. This is the first time this orbital period and the corresponding wind structure are probed in a time-resolved way with a modern X-ray instrument (see Fig.~5 in \citealt{Kretschmar_2021a}).

\subsubsection{X-ray colour evolution with orbital phase}
\label{section:discuss_orbital_evol}

We observe an interesting gradual increase of the hardness ratio between the 3.0--6.0\,keV and 8.0--10.0\,keV energy bands (see third panel of Fig.~\ref{fig:hardness_ratios}). This is more a consequence of a general geometric change in the stellar wind rather than local accretion of clumps. When the wakes are coming through our line of sight (see Fig.~\ref{fig:vela_sketch}), the absorption in the stellar wind increases, preferentially absorbing low-energy photons emitted in the vicinity of the neutron star starting from $\phi_{\mathrm{orb}} \approx 0.44$. In our pulse-by-pulse analysis, this  rise of the absorption column density $N_{\mathrm{H,1}}$ can be directly measured Fig.~\ref{fig:some_param_vs_time_pulse}.

On the other hand, the hardness ratio of high-energy bands is constant (last panel of Fig.~\ref{fig:hardness_ratios}), implying a stable behaviour of the continuum emission from the neutron star. \citet{Martinez_2014}, observed similar behaviour - spectral changes at low energies due to increasing absorption but stable overall source continuum - during their observation, covering eclipse egress and a major flare.

The above is supported by the behaviour of the source on the colour-colour diagram (Sect.~\ref{sec:hardness}) where it describes a nose-shape (Fig.~\ref{fig:nose_plot_0.5-3.0_3.0-6.0_6.0-10.0}). This is typical of the presence of a partial coverer with variable column density in the system \citep[e.g.][in Cyg X-1]{Hirsch_2019, Grinberg_2020}. As would be expected given the onset of the wake, the source evolves along the track with time as indicated by transition from light (early in the observation) to dark green (late in the observation) data points in the figure.

\begin{figure}
    \centering
    \centerline{\includegraphics[trim=0cm 0cm 0cm 0cm, clip=true, width=1.0\linewidth]{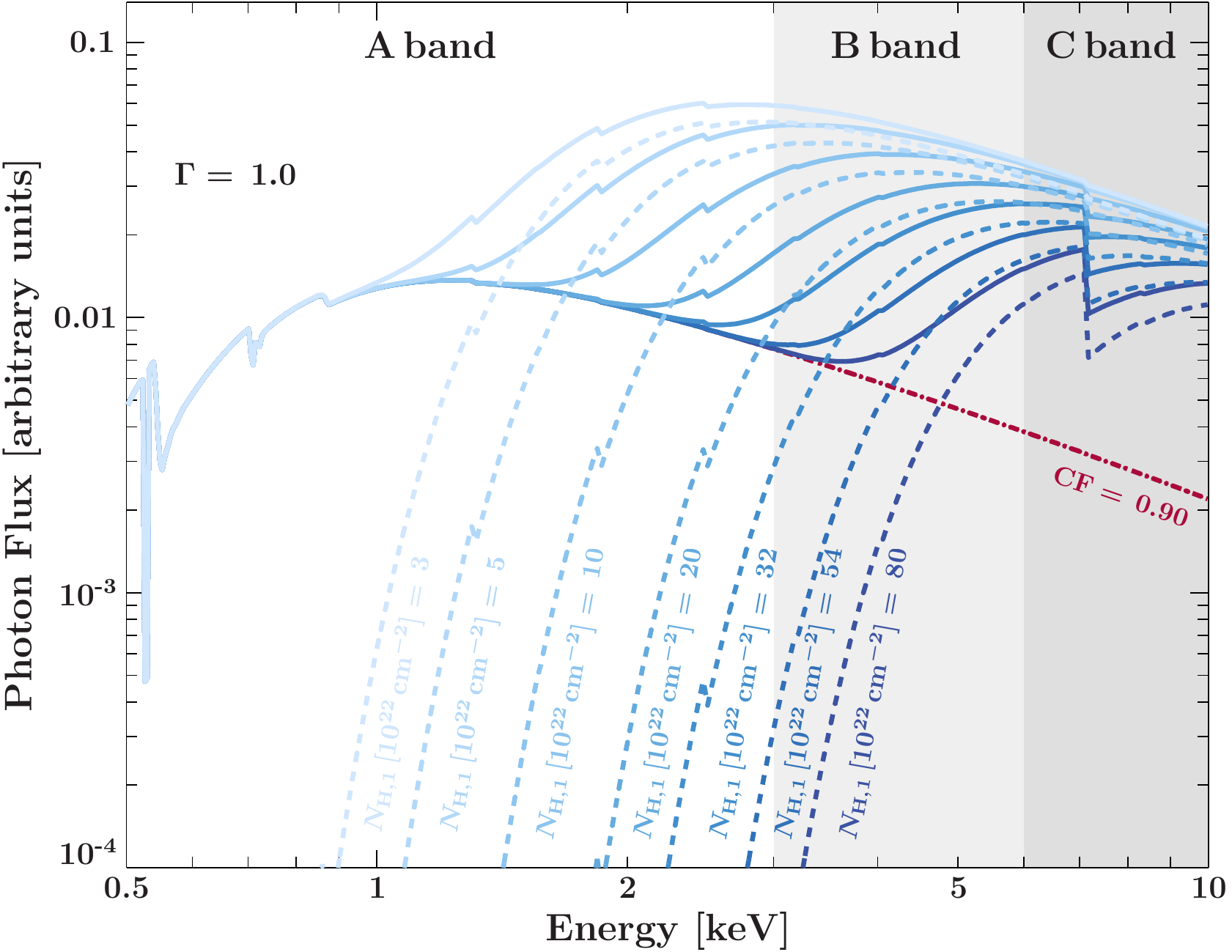}}
    \caption{Effect of increasing absorption on our model from Eq.~\ref{eq:final_model} with photon index $\Gamma = 1.0$ and without emission lines to focus on the evolution of the continuum shape. We assume a covering fraction $\mathrm{CF}$ of 0.9 and a varying absorption column density $N_{\mathrm{H,1}}$ from $3 \times 10^{22}\,\mathrm{cm}^{-2}$ to $80 \times 10^{22}\,\mathrm{cm}^{-2}$ covering the range obtained in Fig.~\ref{fig:some_param_vs_time_pulse}. The shaded grey areas indicate three energy bands of interest: A band from 0.5 to 3\,keV, B band from 3 to 6\,keV and C band from 6 to 10\,keV. The resulting observed spectrum (solid lines) is the sum of the spectrum not covered by the stellar wind (dash-dotted line) and the covered spectrum (dashed lines). See Fig.~3 of \citet{Diez_2022} for an illustrated picture of the partial covering model.}
    \label{fig:variable_nh}
\end{figure}

In Fig.~\ref{fig:variable_nh}, we present how absorption impacts the observed spectrum modelled by Eq.~\ref{eq:final_model}, consisting of a power law continuum with a high-energy cutoff assuming a certain covering fraction $\mathrm{CF}$. In the case of a continuum fully covered by the obscurer ($\mathrm{CF}=1$, dashed lines), the flux ratios in the soft colour (A/B) and in the hard colour (B/C) decrease as $N_{\mathrm{H,1}}$ grows leading to a positive correlation between those ratios for the covered spectrum. 
On the other hand, if we consider a spectrum where this time only a certain fraction $\mathrm{CF}<1$ of the continuum is absorbed by the stellar wind (solid lines), the flux in the A band will remain constant as $N_{\mathrm{H,1}}$ grows after a certain threshold. In the example of a covering fraction of 0.9, this happens at $N_{\mathrm{H,1}} = 54 \times 10^{22}\,\mathrm{cm}^{-2}$ according to expectations from Fig.~\ref{fig:variable_nh}. Simultaneously, the fluxes in the B and C bands continue to decrease together as $N_{\mathrm{H,1}}$ grows. Hence, the softer colour becomes softer as the harder colour does not change leading to the observed nose-shape colour-colour diagram in Fig.~\ref{fig:nose_plot_0.5-3.0_3.0-6.0_6.0-10.0}. This probes the necessity of a partial covering model to describe the data. A higher covering fraction $\mathrm{CF}$ leads to a less elongated shape of the curve.

We use an averaged $\Gamma$ over the values obtained in Fig.~\ref{fig:some_param_vs_time} for our observation. We simulate a grid of colour-colour tracks for varying values of $N_\mathrm{H}$ and covering fraction and include the results in Fig.~\ref{fig:nose_plot_0.5-3.0_3.0-6.0_6.0-10.0}. Data and simulations agree very well, supporting our interpretation.

\subsubsection{Comparison with previous observations and model descriptions}

\begin{figure}[!htb]
    \centering
    \centerline{\includegraphics[clip=true, width=1.0\linewidth]{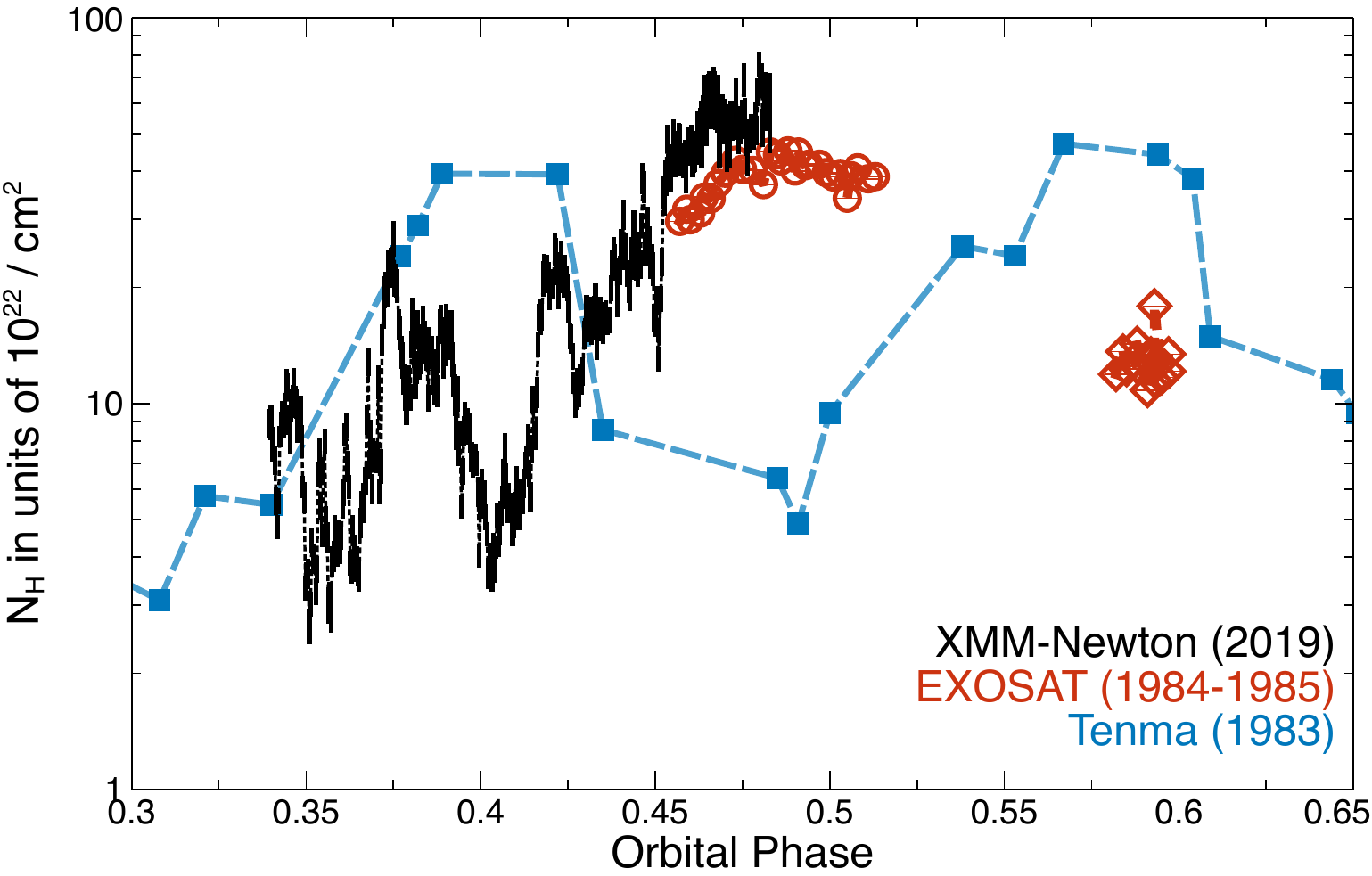}}
    \caption{Comparison of the $N_\mathrm{H}$ values determined in this study with historical measurements taken during individual binary orbits by \textsl{Tenma} \citep{Ohashi_1984} and \textsl{EXOSAT} \citep{Haberl_1990}. Note the overall similar slope of the different rising curves. See text for details.}
    \label{fig:nh_obs}
\end{figure}

While absorption values have been determined by many authors with various different satellites \citep[see][for an overview]{Kretschmar_2021a}, there are few data sets covering a significant range in orbital phase within an individual binary orbit and thus not mixing potential binary orbit-to-orbit variations in wake structures. After correcting for differences in orbital phase definitions in the original papers, in Fig.~\ref{fig:nh_obs} we compare the absorption values we derived with data points from \citet{Ohashi_1984} and \citet{Haberl_1990}. Different spectral models used to fit spectra and to derive $N_\mathrm{H}$ may introduce systematic shifts in the obtained values. Still, the data taken over many years appear to cover a similar range, but sometimes with quite different values at the same orbital phase, as seen already in \citet{Kretschmar_2021a}, indicating that the structures causing these variations are not stable in orbital phase. 
On the other hand, the duration of the overall rising trend from low absorption to a highly absorbed 'plateau' is rather similar in slope -- $N_\mathrm{H}$ values double over a time range of 0.02 in orbital phase or $\sim$15.4~ks -- suggesting that similar larger structures in the wind exist, which may be rather differing in their relative orientation.

\begin{figure}[!htb]
    \centering
    \centerline{\includegraphics[clip=true, width=1.0\linewidth]{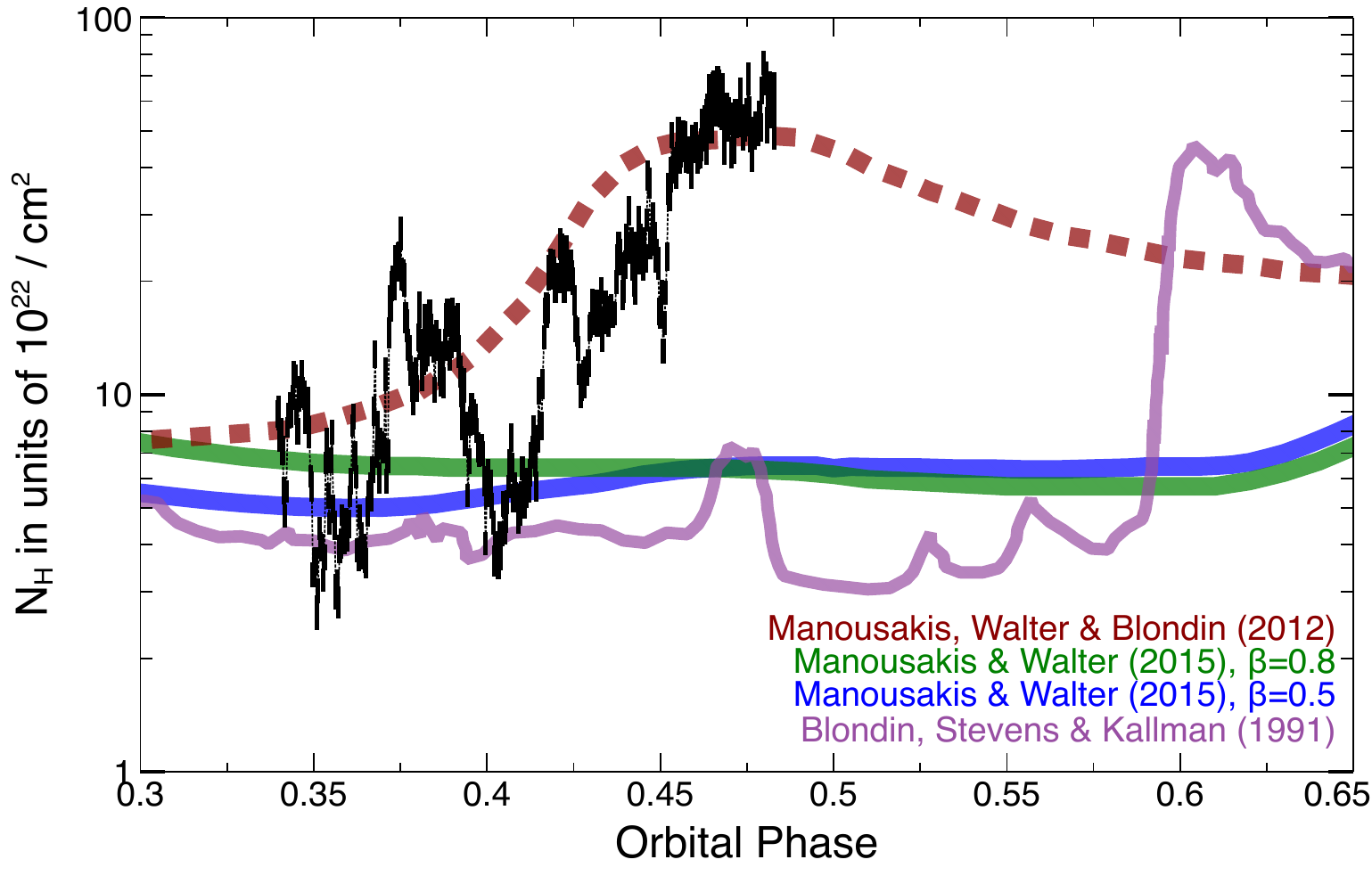}}
    \caption{Comparison of the $N_\mathrm{H}$ values determined in this study with a range of model results for $N_\mathrm{H}$ from hydrodynamic simulations for Vela X-1 or similar, but not identical model systems. See text and Table~\ref{tab:models}}. 
    \label{fig:nh_model}
\end{figure}

\begin{table*}[]
\renewcommand{\arraystretch}{1.1}
    \caption{Parameters used in the model curves shown in Fig.~\ref{fig:nh_model}. All the simulations use circular binary orbits with a fixed distance of the neutron star from the stellar surface. The last row shows parameter ranges found by more recent studies of the Vela~X-1 system (since the year 2000), as compiled by \citet{Kretschmar_2021a}.}
    \label{tab:models}
    \centering
    \begin{small}
     \begin{tabular}{lrrrrrrl}
         System & \multicolumn{2}{c}{donor star} & \multicolumn{2}{c}{stellar wind} & \multicolumn{2}{c}{neutron star} & Ref. and\\
         discussed & \multicolumn{1}{c}{mass ($M_\odot$)}  & \multicolumn{1}{c}{radius ($R_\odot$)} & 
         \multicolumn{1}{c}{mass loss ($M_\odot\,\mathrm{yr}^{-1}$)} & 
         \multicolumn{1}{c}{$v_\mathrm{esc}$ (km s$^{-1}$)} & 
         \multicolumn{1}{c}{$M_\mathrm{X}$ ($M_\odot$)} & 
         \multicolumn{1}{c}{distance ($R_\star$)} & 
         Notes \\
         \hline\hline
         Generic HMXB   & 22   & 35.2 & $5.8\times 10^{-6}$  & 1300 & 1.4  & 1.59 & [B91] \\
         Vela X-1       & 23.1 & 30   & $4\times 10^{-6}$    & 1400 & 1.86 & 1.77 & [MW15b] \\ 
         EXO 1722$-$363 & 15 & 29 & $1\times 10^{-6}$ & 500 & 1.9 & 1.75 &  [MWB12] \\
        \hline
         Vela X-1       & 21--28 & 27--32 & 0.4--$2\times 10^{-6}$ & 380--750 & 1.8--2.1 & 1.56--2 & [K21] \\
         \hline
         \multicolumn{8}{l}{[B91] \citet{Blondin_1991}. Showing the 'full simulation', Fig.~8 in the publication.} \\
        \multicolumn{8}{l}{[MW15b] \citet{ManousakisWalter:2015b}. Stellar parameters from \citet{Manousakis_2015a}.} \\
         \multicolumn{8}{l}{[MWB12] \citet{ManousakisWalterBlondin_2012} } \\
         \multicolumn{8}{l}{[K21] \citet{Kretschmar_2021a}. Distance variation from eccentricity, the binary orbit is very well known.} \\
  \end{tabular}
    \end{small}
\end{table*}

In Fig.~\ref{fig:nh_model} we compare our derived $N_\mathrm{H}$ values with some of the few examples of column densities derived from hydrodynamical model calculations. 
It is important to note, though, that these models start from quite different assumptions (see also Table~\ref{tab:models}) and were not made for direct comparison. First of all, the curves shown for the models of \citet{ManousakisWalter:2015b} and \citet{ManousakisWalterBlondin_2012} are time-averaged values of column density over orbital phase, smoothing out the expected significant $N_\mathrm{H}$ variability from binary orbit to binary orbit, while the result from \citet{Blondin_1991} is taken from the simulation of a single binary orbit in a model including a tidal stream. 
It is evident that neither that last model, nor the relatively low and little varying average column density predicted in  \citet{ManousakisWalter:2015b} shows a marked rise at early orbital phases as found in the data. One of the reasons may be the rather high wind velocities and mass loss rates, in line with earlier studies of Vela~X-1, assumed in these studies, while more modern studies prefer lower wind velocities \citep[see Tab.~7 and Fig.~19 in][]{Kretschmar_2021a}. The visually best matching curve from \citet{ManousakisWalterBlondin_2012} was actually calculated for a different system, EXO~1722$-$363, albeit with quite similar system parameters to Vela X-1 and assuming a relatively slow wind. Updated hydrodynamical simulations accounting for the current best knowledge of orbital and wind parameters of the system, also taking the non-negligible eccentricity into account would be very welcome.

\subsection{Origin and nature of the absorber}

\begin{figure}
    \centering
    \centerline{\includegraphics[trim=0cm 0cm 0cm 0cm, clip=true, width=1.0\linewidth]{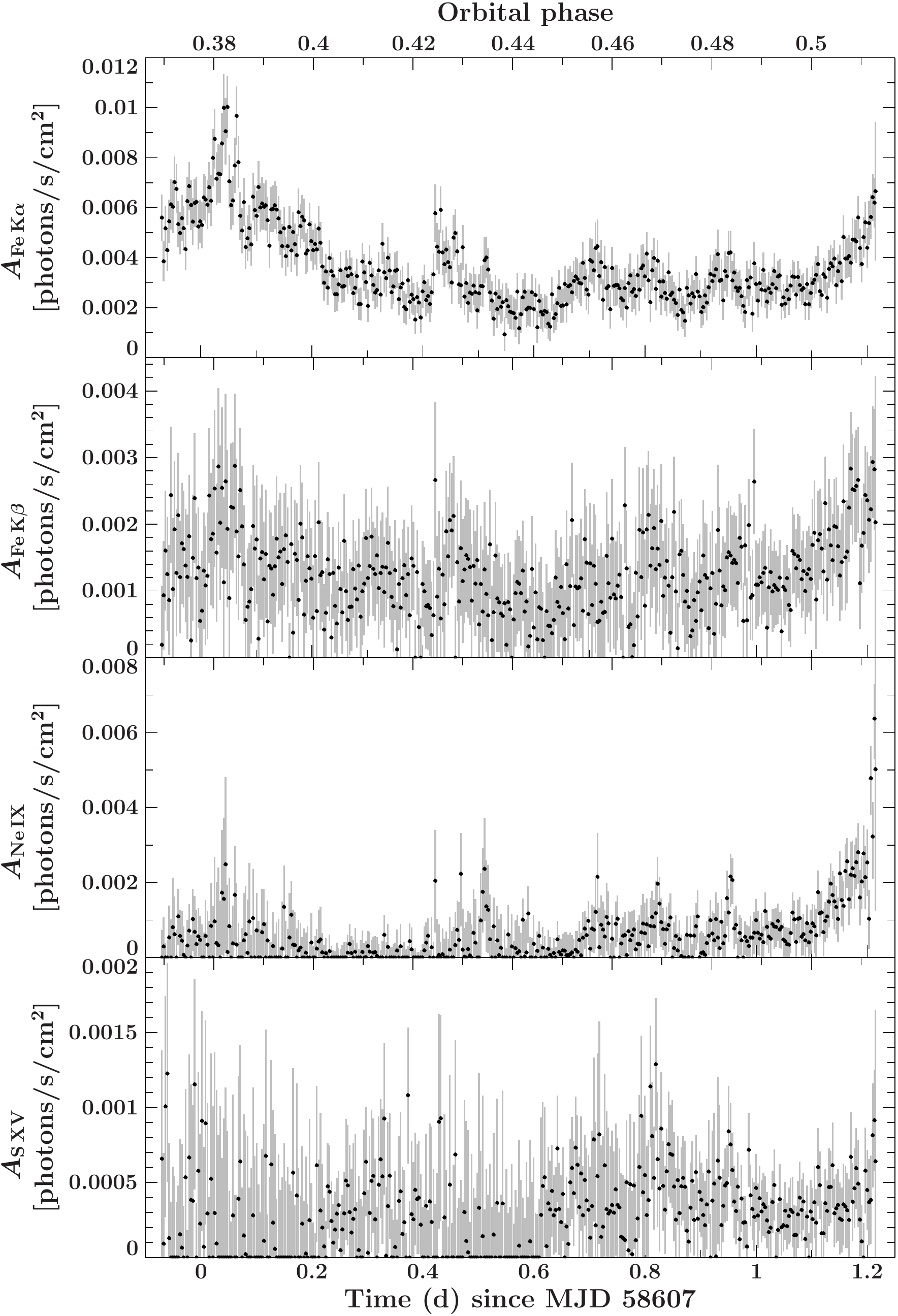}}
    \caption{Fluxes in photons/s/cm$^2$ of some soft lines obtained within the pulse-by-pulse analysis as functions of time, showing also
    the corresponding binary orbital phase. The panels show (from top to bottom): flux of the FeK$\alpha$, the FeK$\beta$, the \NeIX\, and the \SXV\, line.}
    \label{fig:areas_lines}
\end{figure}

The continuum from the neutron star dominates the emission in the spectrum of Vela X-1 until $\phi_{\mathrm{orb}} \approx 0.44$. At this orbital phase, the absorption column density increases together with the strength of soft emission lines between 0.5 and 4\,keV. In Fig.~\ref{fig:areas_lines}, we show the evolution of the flux of some lines with time. We can see in particular the fluxes of the \NeIX\,and \SXV\, are mostly consistent with 0 at the beginning of the observation but they start to increase towards the end revealing the corresponding elements in the spectrum of Vela X-1. The presence of strong lines during heavy absorption from the stellar wind suggests that the absorber is localised and the lines originate from a larger scale in the system as in \citet{Watanabe_2006}. If they originated from the local absorber or close to vicinity of the neutron star, they would be completely absorbed by the stellar wind and would not appear in the resultant spectrum. Another argument in the favour of this statement is that those soft emission lines are also present in the spectrum of Vela X-1 during the eclipse \citep{Sako_1999a} hence when the neutron star and its local absorber are outside the line of sight of the absorber. 

In Fig.~\ref{fig:areas_lines}, the fluxes of the fluorescent $\mathrm{FeK\alpha}$ and $\mathrm{FeK\beta}$ lines are positive throughout the whole observation meaning that those lines are visible at all observed orbital phases here and therefore also originate from a larger scale in the system. It is less evident for the fluorescent \CaIIXII \, $\mathrm{K\alpha}$ line which was more difficult to constrain because of blending with neighbouring lines. 

The \NeIX, \MgXI, \SiXIII \, and \SXV \, complexes are evidences of ionisation of the absorber in the system as those ions only have two remaining electrons on their orbital. Furthermore, the presence of \OVIII, \NeX, \MgXII, \SiXIV \, and \SXVI \, \Lya \, lines also indicates ionisation in the system as they are emitted when the last atomic electron transitions from an $n = 2$ orbital to the ground state. According to \citet{Amato_2021a}, the warm photoionised wind of the companion star and smaller cooler regions or clumps of gas can explain the simultaneous contribution of H- and He-like emission lines and fluorescent lines of near-neutral ions. Comparing \chandra/HETGS data of Vela X-1 with simulations of propagation of X-ray photons in a smooth and undisturbed wind, \citet{Watanabe_2006} stated that fluorescent lines originate from reflection of the stellar photosphere in the extended stellar wind or simply from the accretion wake. Additionally, they observed brighter soft emission lines at $\phi_{\mathrm{orb}} = 0.5$ than during the eclipse. This indicates a higher production of X-ray line emission caused by highly ionised ions in a region between the neutron star and its massive stellar companion, which is occulted during eclipse. This is however difficult to confirm with our data and further studies at higher spectral resolution are necessary.

\subsection{Short-term absorption variability}

\subsubsection{Search for characteristic timescales in absorption variability}

\begin{figure}
    \centering
    \includegraphics[width=1.0\linewidth]{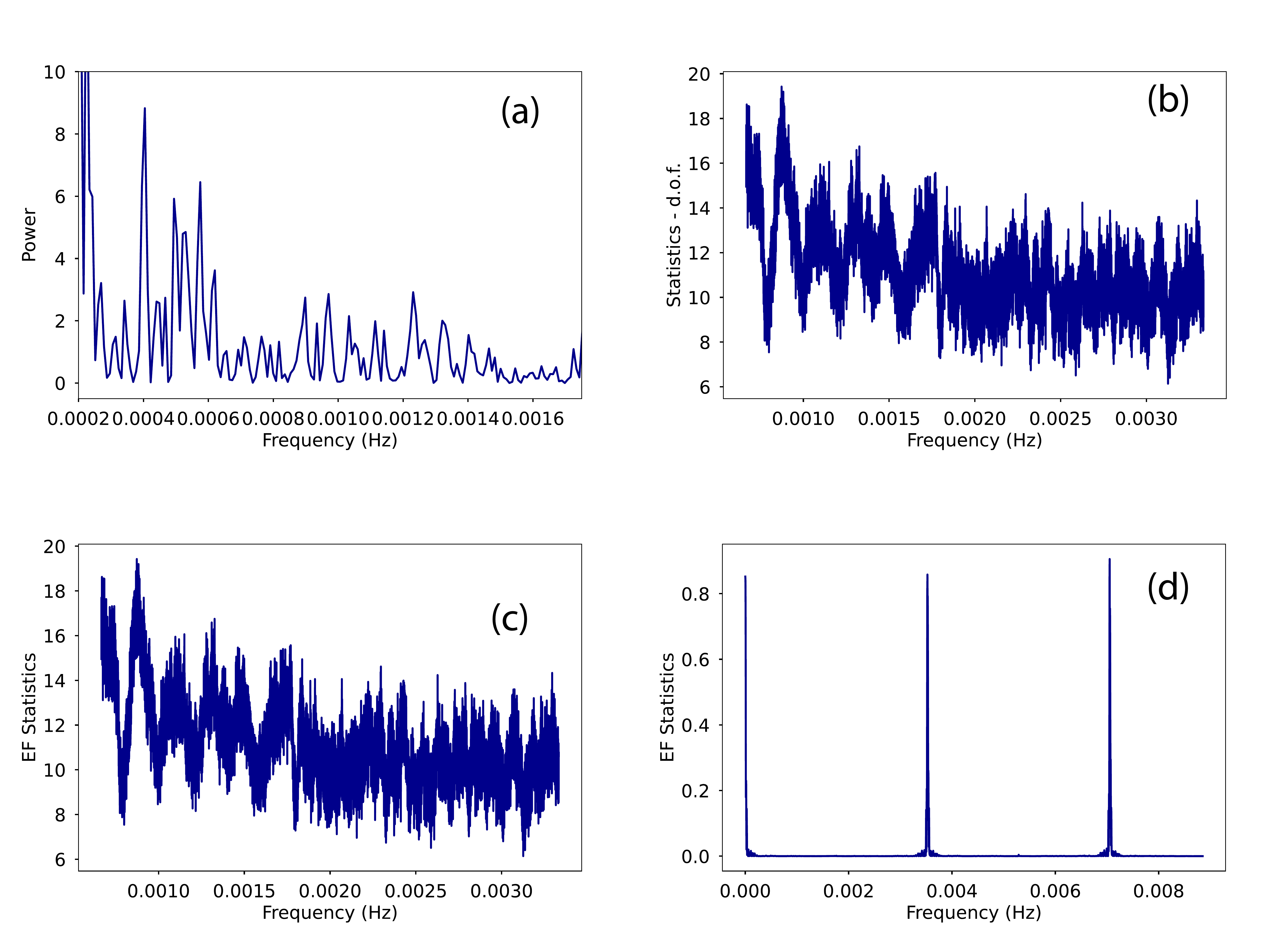}
    \caption{Timing analysis of $N_{\mathrm{H}}$ evolution: (a) Stingray average power spectrum; (b) Stingray z-square function; (c) Stingray chi-square function; (d) Astropy v5.1 Lomb-Scargle Periodogram.}
    \label{fig:timing_Nh}
\end{figure}

Theoretical predictions show that variability in a clumpy material can result in typical variability timescales that will depend on the properties of the structures in the stellar wind and/or in the tidal streams in correlation with the orbital parameters of the system \citep{El_Mellah_2020a}.
In this section, we thus search for possible indications of such a timescale. To perform this analysis, we have used the \textit{Stingray: A Modern Python Library for Spectral Timing} ~\citep{stingray_software,Huppenkothen_2019,Huppenkothen_2019_soft} Python library. Our pseudo-light curve of $N_{\mathrm{H}}$ contains 392 bins with a binning size equal to the pulse period of the neutron star ($\sim$283 s). We have used different techniques implemented in Stingray which are: 

\begin{itemize}

\item Average power spectrum using Leahy normalisation and dividing our data into 11 subsets. This technique allows the search for periodicities in the frequency range between 0.0002 to 0.00175 Hz (see panel (a) of Fig.~\ref{fig:timing_Nh}).

\item z-search and chi-search techniques to search for periodicity in the frequency range between 0.006 to 0.0033 (see  panels (b) and (c) of Fig.~\ref{fig:timing_Nh} respectively).

\end{itemize}

Neither of the techniques detects a significant periodic or quasi-periodic signal in the evolution of $N_{\mathrm{H}}$ in the range of frequencies accessible with our data. The simulations of \citet{El_Mellah_2020a} predict a clear signal in the autocorrelation function equivalent to a cutoff in the power spectrum. Given the quality of our data, the presence of such a feature cannot be assessed. Still, the above presented is, to our knowledge, the first absorption power spectrum calculated for Vela X-1 in an attempt to obtain such timescales.

To expand the frequency search range to larger values, we have also performed a Lomb-Scargle Periodogram using the routine of Astropy v5.1 Timeseries software \citep{astropy:2022} (see panel (d) of Fig.~\ref{fig:timing_Nh}). This technique only finds a possible period the binning size of our database and their following harmonics but no another possible periodicity. It does not allow the shape of the power spectrum to be assessed. This search could be enhanced with an extended sample of the $N_{\mathrm{H}}$ along one or more binary orbits.

\subsubsection{Absorption during flares}

During the first two flaring episodes at $\phi_{\mathrm{orb}} \approx 0.38$ and $\phi_{\mathrm{orb}} \approx 0.43$, we saw in the second panel of Fig.~\ref{fig:hardness_ratios} that the hardness ratio between the continuum band and the softest band decreases dramatically indicating a softening of the underlying spectrum. This is also confirmed in Fig.~\ref{fig:some_param_vs_time} with the local decreases of photon index during flaring episodes. Therefore, more low-energy photons are detected compared to before the flaring episodes. If more low-energy photons reach the detector plane, it suggests that there is less material on the line of sight of the observer (otherwise they would have been absorbed by the wind) so less absorption from the stellar wind. This has been confirmed by our time-resolved spectral analysis in Fig.~\ref{fig:some_param_vs_time} and in Fig.~\ref{fig:some_param_vs_time_pulse}, where the absorption column density of the stellar wind $N_{\mathrm{H,1}}$ reaches its minimum during those two flaring episodes. 

This is also visible during later short flares with strong but brief softening of the spectral shape together with local minima of $N_{\mathrm{H,1}}$ and $\mathrm{CF}$. These short timescale events could be associated with accretion of clumps in the vicinity of the neutron star as already suggested in \citet{Martinez_2014} and \citet{Diez_2022} for Vela X-1. As material falls onto the surface of the neutron star through the accretion column, photons are produced through bremsstrahlung and cyclotron emission. Those photons are then up-scattered through inverse Compton and more X-rays are produced. The more material falls into the neutron star, the more the temperature increases favouring interactions and X-ray production. Clumps just passing in front of the source on the line of sight of the observer could explain the local maxima in $N_{\mathrm{H,1}}$ that are not happening simultaneously with flux changes.

\section{Summary and outlook}
\label{section:summary}

We have analysed simultaneous \xmmnewton and \nustar data of Vela X-1 covering a broad X-ray range at orbital phase $\sim$0.36--0.52. For the spectral modelling, we used our partial covering model first described in \citet{Diez_2022}. Thanks to the hard X-ray coverage permitted by \nustar and our results from previous work, we were able to constrain the continuum in order to focus on the absorption variability at lower energies with \xmmnewton EPIC-pn for this work.

This is the first time that we have such a high-time-resolution absorption study of Vela X-1 on a broad X-ray range from 0.5 to 78\,keV. We have traced the onset of the wakes characterised by a rise of the absorption column density $N_{\mathrm{H,1}}$ starting at orbital phase $\sim$0.44 as well as local absorption variability due to accretion of clumps. The slope of the $N_{\mathrm{H,1}}$ rise is comparable and similar to previous observations, albeit with an orbital phase lag indicating similar large scale structures in the wind but with different orientation at different times of observation. We also compared our data with simulations from previous works in the literature but no strong match between observations and theoretical models could be found. This reflects the necessity for further and updated hydrodynamical simulations accounting for the latest orbital parameters values (for example: eccentricity > 0) and complexity of the wind parameters (such as wind velocities, mass loss rates).

Through high-resolution spectroscopy of the multiple fluorescent lines present in Vela X-1, we performed the X-ray photography of the material in the system. The evidence of those lines at different absorption phases suggests sources of emission from local absorber to large-scale structures. This analysis also revealed strong photoionisation of the wind with the presence of highly ionised elements such as \Lya\,lines of O, Ne, Mg, Si and S. However, these results have to be considered with caution as the \xmmnewton EPIC-pn energy resolution is not sufficient to perform an accurate spectral analysis of individual lines and blending with neighbouring elements can happen. This aspect is beyond the scope of this paper and the analysis of the simultaneous \xmmnewton RGS data in a future work is needed to disentangle this. Moreover, we used a neutral absorber to describe the absorption from the wind but the use of \texttt{warmabs}\footnote{\url{https://heasarc.gsfc.nasa.gov/xstar/docs/html/node102.html}} for warm absorbers and photoionised emitters as well as a higher resolution instrument (such as \chandra/HETGS) would be better suited. The upcoming \xrism and \athena \citep{Athena} are of utmost importance for such a study as high-resolution spectral analysis is one of their main science goals \citep{XRISM_2020}.

\begin{acknowledgements} This work has been partially funded by the  Bundesministerium f\"ur Wirtschaft und Energie under the Deutsches Zentrum f\"ur Luft- und Raumfahrt Grants 50~OR~1915.
The research leading to these results has received funding from the
European Union’s Horizon 2020 Programme under the AHEAD2020 project (grant agreement n. 871158) and from the ESA Archival Research Visitor Programme.
SMN acknowledges funding under project PID2021-122955OB-C41 funded by MCIN/AEI/10.13039/501100011033 and by
‘ERDF A way of making Europe’.
RA acknowledges support by the CNES.
Work at LLNL was performed under the auspices of the US Department of Energy under contract No. DE-AC52-07NA27344 and supported through NASA grants to LLNL.
This work has made use of (1) the Interactive Spectral Interpretation System (ISIS) maintained by Chandra X-ray Center group at MIT; (2) the \nustar Data Analysis Software (NuSTARDAS) jointly developed by the ASI Science
Data Center (ASDC, Italy) and the California Institute of Technology (Caltech, USA); (3) the ISIS functions (\texttt{isisscripts})\footnote{\url{http://www.sternwarte.uni-erlangen.de/isis/}\label{footnote:isisscripts}} provided by ECAP/Remeis observatory and MIT; (4) NASA's
Astrophysics Data System Bibliographic Service (ADS); (5) the Users Guide to the XMM-Newton Science Analysis System, Issue 17.0, 2022 (ESA: XMM-Newton SOC). We thank John E. Davis for the development of the
\texttt{slxfig}\footnote{\url{http://www.jedsoft.org/fun/slxfig/}\label{footnote:jedsoft}}
module used to prepare most of the figures in this work. Others were created with the Veusz\footnote{\url{https://veusz.github.io/}} package.
WebPlotDigitizer\footnote{\url{https://automeris.io/WebPlotDigitizer/}}
(\textcopyright\ Ankit Rohatgi)
has been used to digitise data from figures in older publications.

\end{acknowledgements}

\bibliographystyle{aa}
\bibliography{references}

\begin{appendix}

\section{Calibration of \xmmnewton EPIC-pn timing mode}
\label{appendix:calib_title}

In this section, we discuss the calibration of the \xmmnewton EPIC-pn timing mode and the tests we performed to justify our choice of calibration and data extraction for this work. We will discuss our tests and their implications to conclude with the possible caveats.

\subsection{Test for RDCTI and RDPHA correction in the iron line region}
\label{appendix:iron_line_calib_epproc}

When processing the ODFs to obtain calibrated and concatenated event lists to later generate scientific products, one has to use the \texttt{epproc} task. The default calibration of this task for timing mode data\footref{footnote:epproc} uses \texttt{withrdpha='Y', withxrlcorrection='Y', runepreject='Y', runepfast='N'}; \texttt{Y} and \texttt{N} standing for \texttt{YES} and \texttt{NO} respectively. 

The Rate Dependent PHA (RDPHA) correction was introduced with SASv13 as a more robust method than the Rate Dependent CTI (RDCTI) correction to rectify for count-rate dependent effect on the energy scale of EPIC-pn exposures in timing mode\footnote{see the CCF Release Note 0312 (Guainazzi M., 2014a, XMM-CCF-REL-0312) \url{https://xmmweb.esac.esa.int/docs/documents/CAL-SRN-0312-1-4.pdf} and the Science Operations Team Calibration Technical Note 0083 (Guainazzi M., et al. 2014b, XMM-SOC-CAL-TN-0083) \url{https://xmmweb.esac.esa.int/docs/documents/CAL-TN-0083.pdf}}. Thus, the task \texttt{epfast}, which applies the RDCTI correction, does not run on data which has been already corrected with the RDPHA correction, and vice-versa. This explains why \texttt{runepfast} is set to \texttt{'N'} since \texttt{withrdpha} is set to \texttt{'Y'} by default in the timing mode.

\begin{figure}
    \centering
    \centerline{\includegraphics[trim=0cm 0cm 0cm 0cm, clip=true, width=1.0\linewidth]{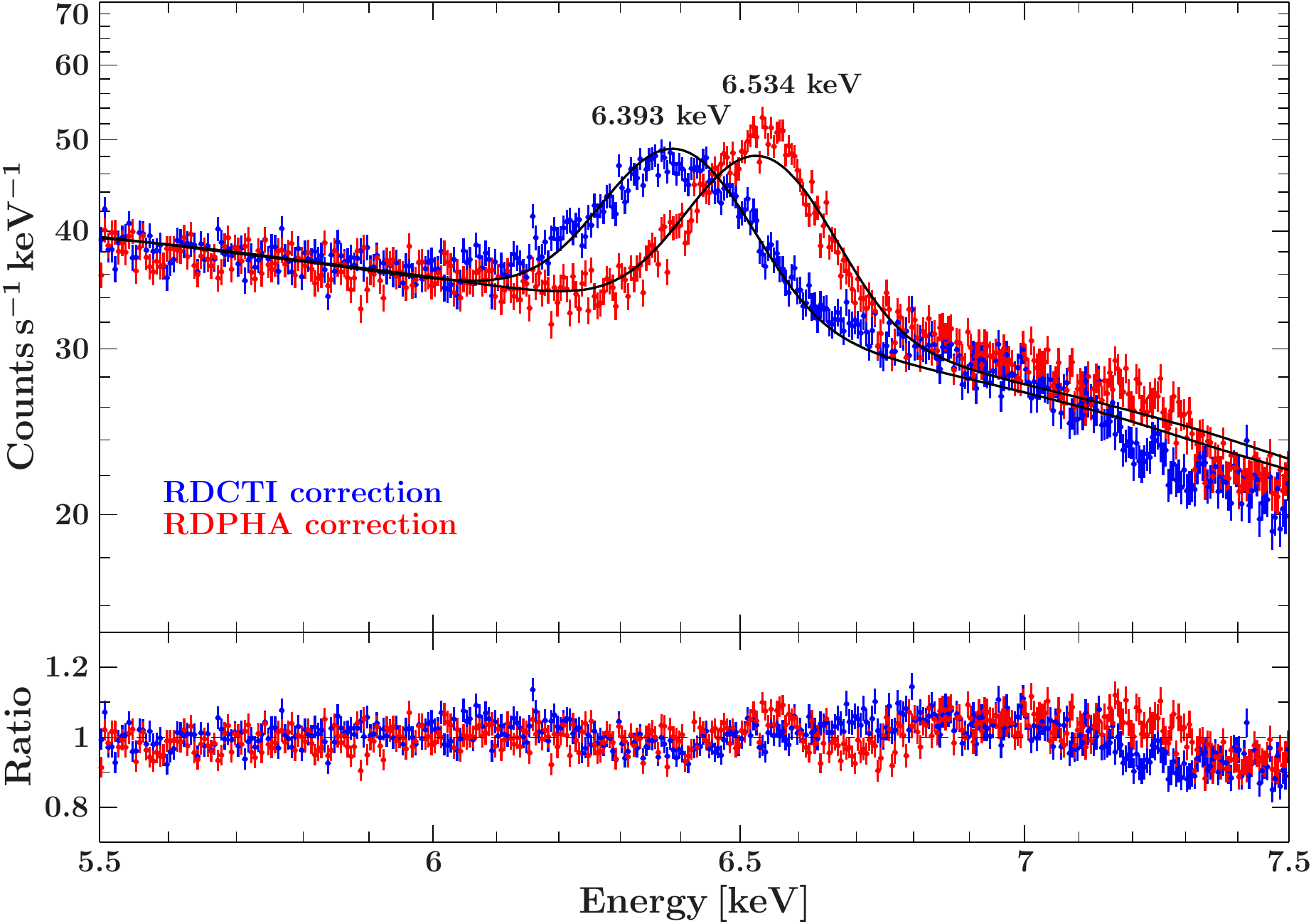}}
    \caption{Example of a spectrum in the iron line region generated with different calibrations. The red spectrum corresponds to events generated applying the RDPHA correction, while the blue spectrum corresponds to the RDCTI correction.}
    \label{fig:iron_line_calib_epproc}
\end{figure}

However, when extracting the \xmmnewton EPIC-pn spectra for this Vela X-1 observation with the timing mode default RDPHA correction, we obtained higher energies than expected for the line features in the 0.5--10\,keV energy range covered by EPIC-pn. It is particularly visible for the fluorescent emission line associated with $\mathrm{FeK\alpha}$ as it is the most prominent emission feature in the spectrum of Vela X-1. We present in Fig.~{\ref{fig:iron_line_calib_epproc}} an example of a spectrum with the RDPHA correction in the iron line region fitted with a simple power law and a Gaussian component. The $\mathrm{FeK\alpha}$ line is found at more than 6.53\,keV, while it is expected to be located around $\sim$6.4\,keV according to our results of the simultaneous \nustar observation \citep{Diez_2022} or in the spectrum of Vela X-1 in general \citep[see e.g.][]{Goldstein_2004,Watanabe_2006,Gimenez-Garcia_2016}. The energy of the iron line also depends on how much the observation is affected by pile-up, this will be discussed in the following section. 

In order to perform a sanity check of the RDPHA correction, we decide to reverse back to the RDCTI correction (\texttt{epfast}) as it was usually done in timing mode before the RDPHA correction got released in SASv13 and later versions \citep{Martinez_2014}. We therefore tested \texttt{epproc} running the \texttt{epfast} task, so using the parameters \texttt{withdefaultcal='N', withrdpha='N', withxrlcorrection='Y', runepreject='Y', runepfast='Y'}. This setting corresponds to the default calibration of the burst mode\footref{footnote:epproc}. This results in an iron line around 6.34\,keV as presented in Fig.~\ref{appendix:iron_line_calib_epproc}, which is more consistent with what we expected. The shift we obtained between both calibrations is of $\sim$140\,eV which may impact the quality of our results. It was also seen in the instrumental line features of the detector such as the gold edge region around $\sim$2.2\,keV. The same behaviour has been obtained in \citet{Pintore_2014} in the spectra of the accreting neutron star GX 13+1 where they found a shift of 360\,eV in the iron line between the RDCTI and RDPHA corrections.

\subsection{Test for pile-up in the iron line region}
\label{appendix:iron_line_pile_up}

\begin{figure}
    \centering
    \centerline{\includegraphics[trim=0cm 0cm 0cm 0cm, clip=true, width=1.0\linewidth]{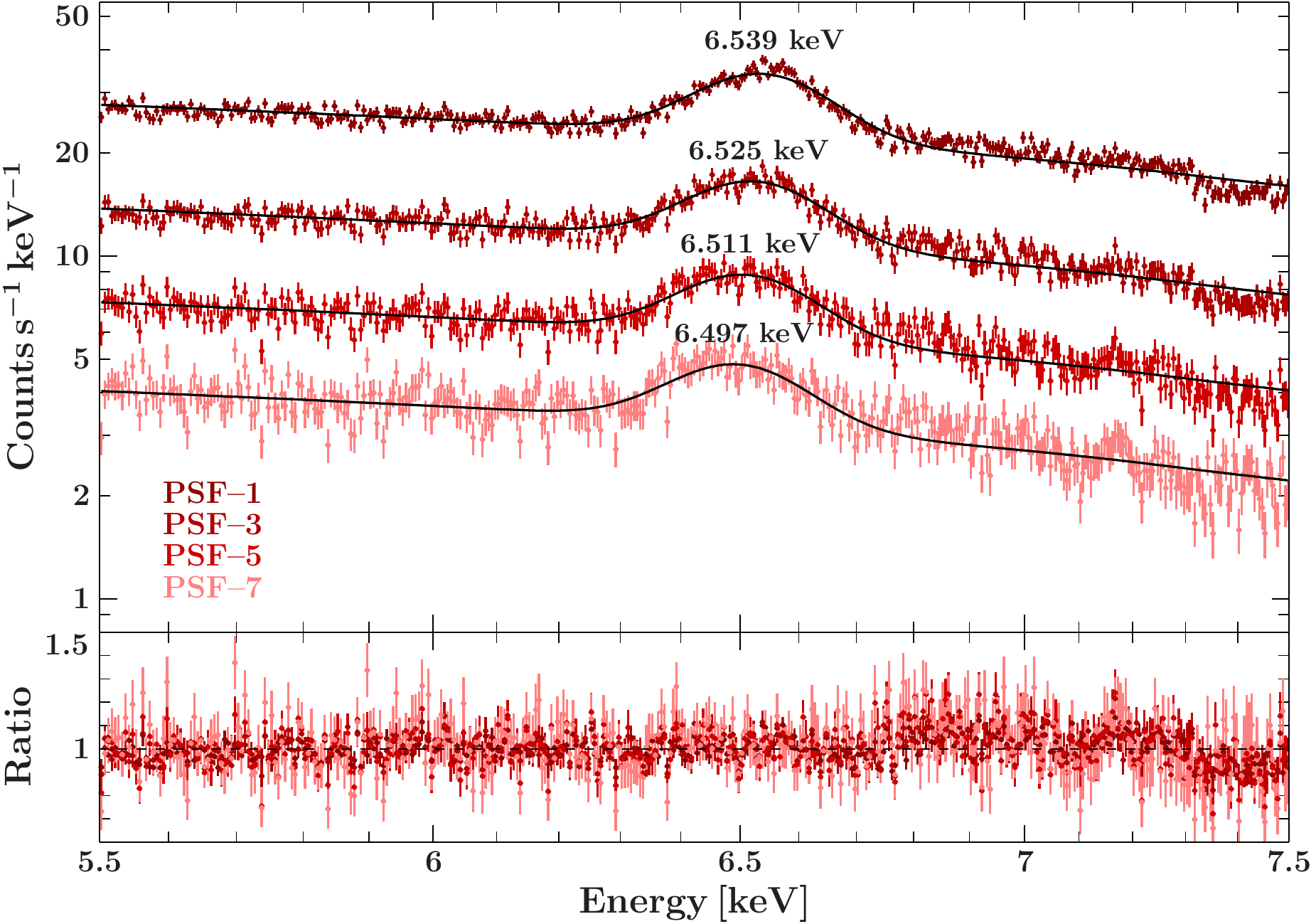}}
    \caption{Example of a spectrum with RDPHA correction as in Fig.~\ref{fig:iron_line_calib_epproc} removing 1 (PSF--1), 3 (PSF--3), 5 (PSF--5) and 7 (PSF--7) columns from the centre of the PSF.}
    \label{fig:iron_line_pile_up_defaultcal}
\end{figure}

\begin{figure}
    \centering
    \centerline{\includegraphics[trim=0cm 0cm 0cm 0cm, clip=true, width=1.0\linewidth]{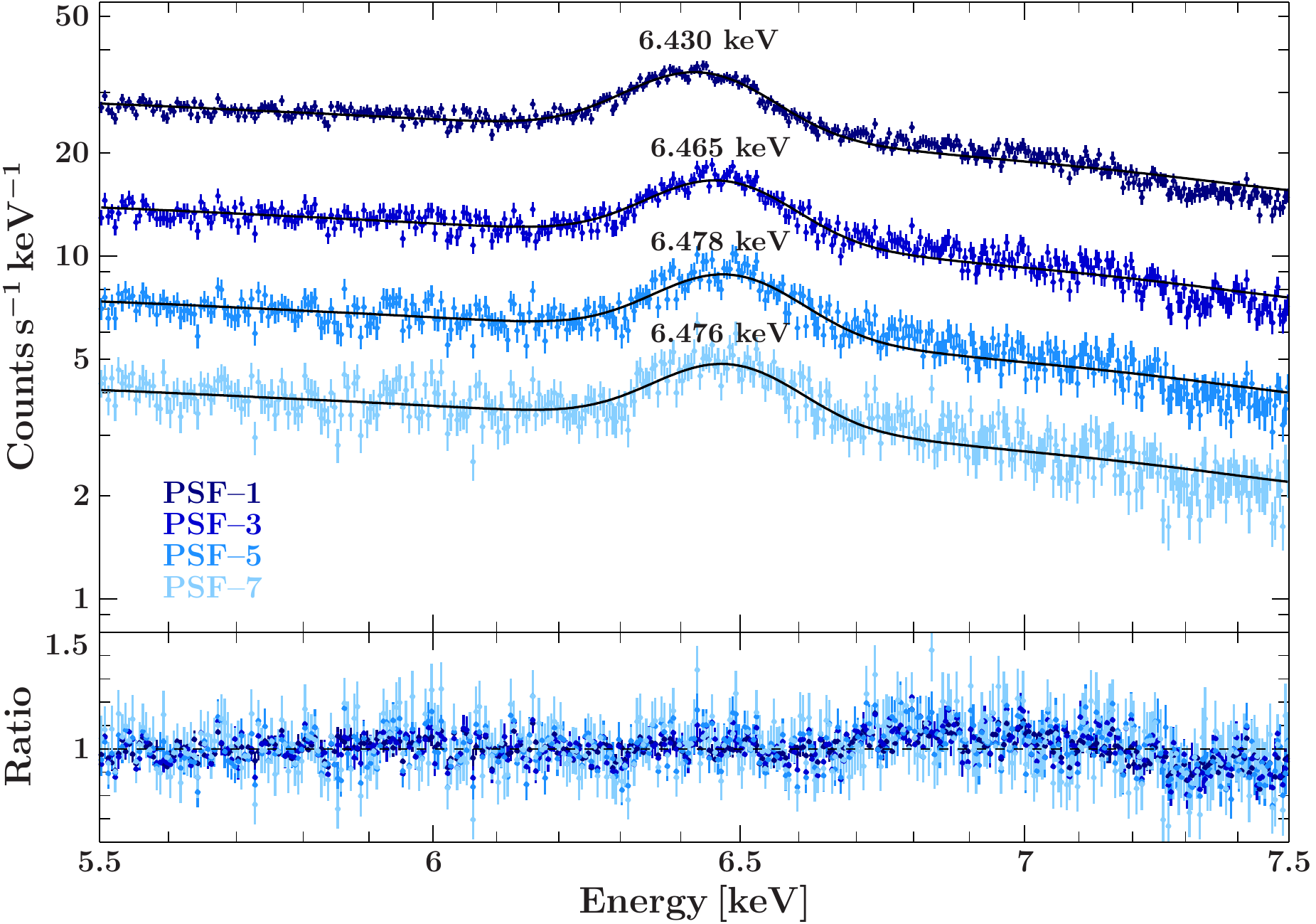}}
    \caption{Example of a spectrum with RDCTI correction as in Fig.~\ref{fig:iron_line_calib_epproc} removing 1 (PSF--1), 3 (PSF--3), 5 (PSF--5) and 7 (PSF--7) columns from the centre of the PSF.}
    \label{fig:iron_line_pile_up_nodefaultcal_epfast}
\end{figure}

As mentioned in Sect.~\ref{section:data_reduc}, our \xmmnewton EPIC-pn observation of Vela X-1 is deeply affected by pile-up. In addition with the \texttt{epatplot} task, we evaluate the pixel columns most affected by pile-up by performing a test in the iron line region. We extract spectra ignoring 1, 3, 5 and 7 columns from the PSF centre (named PSF--\#) and then compare the position of the iron line between different extractions. We present example spectra as in Fig.~\ref{fig:iron_line_calib_epproc} using RDPHA correction (see Fig.~\ref{fig:iron_line_pile_up_defaultcal}) and RDCTI correction (see Fig.~\ref{fig:iron_line_pile_up_nodefaultcal_epfast}) and fitted with a power law and Gaussian component to model the iron line.

We can notice that with the RDPHA correction, the more the centremost columns are removed, the lower the energy of the iron line is starting from $\sim$6.54\,keV for 1 column removed to $\sim$6.50\,keV for 7 ignored columns. However, removing 7 columns from the centre of the PSF is ignoring almost the entire available signal decreasing the signal to noise ratio. Moreover, the iron line energy is still too high compared to what we expect for this feature as discussed in Sect.~\ref{appendix:iron_line_calib_epproc}. On the contrary, using the RDCTI correction, we obtain the opposite behaviour with an increase of the iron line energy together with the number of centre columns removed from the PSF. In order to have a good balance between keeping enough signal and having a consistent iron line energy with previous results for Vela X-1, we decide to apply the RDCTI correction and to remove the three centremost columns from the PSF (PSF--3) for this work. Even if we apply those corrections, the energy of the iron line ($\sim$6.46\,keV) is still higher than what we obtained for the simultaneous \nustar observation so discrepancies between both instruments are expected for the combined spectral analysis that will be performed in this work in Sect.~\ref{subsection:spectral_analysis_orbit_XMM_NuSTAR}.

\subsection{Cross-instrumental issues}
\label{appendix:cross-instrumental_issues}

\begin{figure}
    \centering
    \centerline{\includegraphics[trim=0cm 0cm 0cm 0cm, clip=true, width=1.0\linewidth]{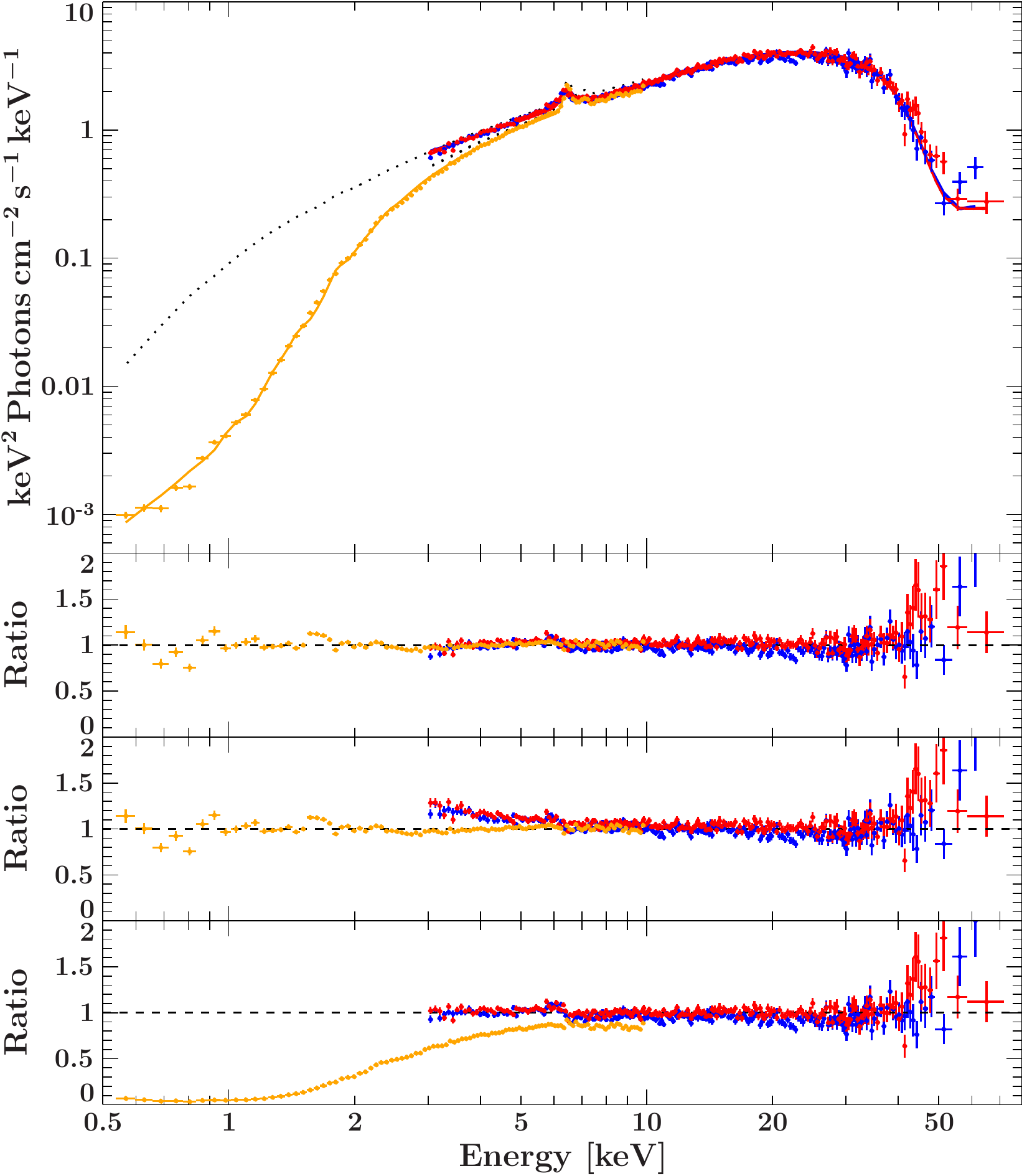}}
    \caption{Example of an unfolded spectrum taken with \xmmnewton EPIC-pn (orange), \nustar FPMA (red) and FPMB (blue) during the fourth \nustar-orbit. Solid lines show the best-fit model with independent $\mathrm{CF_{\textsl{XMM}}}$ and $\mathrm{CF_{\nustar}}$, dotted lines account for best-fit model with tied covering fractions. First residual panel: best-fit model with independent covering fractions. Second residual panel: best-fit model with $\mathrm{CF_{\nustar}}$ tied to $\mathrm{CF_{\textsl{XMM}}}$. Last residual panel: best-fit model with $\mathrm{CF_{\textsl{XMM}}}$ tied to $\mathrm{CF_{\nustar}}$. A strong soft excess in the \nustar data at $\sim$3\,keV comparatively with \xmmnewton data is visible.
    }
    \label{fig:ex_spec_XMM_NuSTAR}
\end{figure}

In comparison with \xmmnewton EPIC-pn data, we observe a strong soft excess in the \nustar data (see Fig.~\ref{fig:ex_spec_XMM_NuSTAR}). A possible explanation can be the dust scattering effect. In this scenario, we suppose dense interstellar clouds located between the observer and the source. Those clouds will deviate the trajectory of low-energy photons that were not supposed to be observed and will be scattered towards the observer producing a soft excess flux at low energy in the spectrum of the source. This scattering leads to the formation of the scattering halo as observed in Circinus X-1 in \citet{Heinz_2015}. To check for this phenomenon, we extracted source regions of different sizes in the \nustar data from 10 arcsec to 50 arcsec by step of 10 arcsec. The smallest source regions account for the maximum of the PSF, though with a possible loss of information from the source. The largest source regions account for the PSF and outside wings, so where the dust scattering effect is visible and more low-energy photons are present. When comparing the resultant spectra, we could not see any difference in terms of spectral shape meaning the soft excess at low energy is always present. Hence, we exclude the dust scattering effect as a cause for the observed soft excess in our \nustar data. We also checked for contamination sources in the extracted source region of both instruments but none could be found. As we are unable to find a plausible physical explanation to this phenomenon, we tested our model by tying $\mathrm{CF_{\nustar}}$ to $\mathrm{CF_{\textsl{XMM}}}$. In the second residual panel of Fig.~\ref{fig:ex_spec_XMM_NuSTAR}, we observe that the \xmmnewton EPIC-pn data are well described albeit the residuals increase drastically for \nustar around 3\,keV. Reciprocally, we tied $\mathrm{CF_{\textsl{XMM}}}$ to $\mathrm{CF_{\nustar}}$ and, as expected, the \nustar data are well constrained. However, there is a discrepancy up to a factor of 10 between the model and the \xmmnewton EPIC-pn data at 0.5\,keV as shown on the last residual panel of Fig.~\ref{fig:ex_spec_XMM_NuSTAR}. None of the datasets is correctly described by an averaged $\mathrm{CF}$. We conclude that the $\mathrm{CF_{\textsl{XMM}}}$ is the most reliable value for the covering fraction and that the problem may be due to remaining calibration effects from \nustar at low energies. For the rest of this work, we will assume two independent covering fractions for each instrument as it is the best way to empirically compensate the observed difference (see the first residual panel of Fig.~\ref{fig:ex_spec_XMM_NuSTAR}). In \citet{Tsygankov_2019}, an offset between \swift/XRT and \nustar has also been reported. The \swift/XRT normalisation was about 1.3 times lower than \nustar FPMA/FPMB possibly due to the fact that their \nustar and \swift observations were not strictly simultaneous.

\end{appendix}

\end{document}